\documentclass[12pt,preprint]{aastex}

\usepackage{multirow}
\def\hst{{\it HST}}

\shorttitle{The M31 Velocity Vector I}
\shortauthors{Sohn et al.}

\begin{document}

\title{The M31 Velocity Vector. \\ 
I. Hubble Space Telescope Proper Motion Measurements}

\author{
Sangmo Tony Sohn, Jay Anderson, and Roeland P. van der Marel}
\affil{Space Telescope Science Institute, 
       3700 San Martin Drive, Baltimore, MD 21218}
\email{tsohn@stsci.edu}

\begin{abstract}
We present the first proper motion measurements for the galaxy M31.
We obtained new $V$-band imaging data with the {\it Hubble Space
Telescope} ACS/WFC and the WFC3/UVIS instruments of three fields: 
a spheroid field near the minor axis, an outer disk field along the 
major axis, and a field on the Giant Southern Stream. The data 
provide 5--7 year time baselines with respect to pre-existing deep 
first-epoch observations of the same fields. We measure the positions 
of thousands of M31 stars and hundreds of compact background galaxies 
in each field. High accuracy and robustness is achieved by building 
and fitting a unique template for each individual object. The average 
proper motion for each field is obtained from the average motion of 
the M31 stars between the epochs with respect to the background 
galaxies. For the three fields, the observed proper motions 
$(\mu_{W},\mu_{N})$ are, in units of mas yr$^{-1}$, 
$(-0.0458, -0.0376) \pm (0.0165, 0.0154)$, $(-0.0533, -0.0104) \pm 
(0.0246, 0.0244)$, and $(-0.0179,-0.0357) \pm (0.0278, 0.0272)$, 
respectively. The ability to average over large numbers of objects 
and over the three fields yields a final displacement accuracy of a 
few thousandths of a pixel, corresponding to only 
$12\ {\mu}$as yr$^{-1}$ . This is comparable to what has been 
achieved for other Local Group galaxies using VLBA observations of 
water masers. Potential systematic errors are controlled by an 
analysis strategy that corrects for detector charge transfer 
inefficiency, and spatially- and time-dependent geometric distortion 
and point-spread-function variations. The robustness of the 
proper-motion measurements and uncertainties are supported by the 
fact that data from different instruments, taken at different times 
and with different telescope orientations, as well as measurements of 
different fields, all yield statistically consistent results. 
Papers~II and~III of this series explore the implications of the new 
measurements for our understanding of the history, future, and mass 
of the Local Group.
\end{abstract}


\section{Introduction}\label{sec:Intro}

At a distance of $\sim770$ kpc \citep[e.g.,][]{fre90}, the Andromeda 
galaxy M31 is the nearest giant spiral to the Milky Way. Together, 
these two galaxies dominate the mass and dynamics of the Local Group. 
It is therefore of tremendous interest to determine the velocity 
vector of M31 with respect to the Milky Way. While the line-of-sight 
velocity of M31 is well-known from Doppler measurements, a 
determination of its proper motion (PM) has so far remained elusive.

The PM of M31 has been sought for almost a century
\citep[e.g.,][]{bar17}, but {\it very high} astrometric accuracy is
required to accomplish this. At the distance of M31, 100 km s$^{-1}$
corresponds to 0.027 mas yr$^{-1}$. This is well beyond the accuracy 
limit of current ground-based optical observations even for long 
time baselines \citep[e.g., $> 40$ year; e.g.,][]{vie10}. The two 
most accurate ($\sim 0.01$ mas yr$^{-1}$) PM measurements in the 
Local Group come from studies of water masers in M33 \citep{bru05} 
and IC 10 \citep{bru07}. However, water masers in M31 have only 
recently been discovered \citep{dar11}, and it will take time until 
a sufficient baseline has been established to enable PM measurements.

Theoretical modeling of the local Universe has suggested that the
Galactocentric rest-frame velocity of M31 must have a tangential
component $V_{\rm tan} \leq 200$ km s$^{-1}$. However, within this 
limit many different orbits are still possible \citep{pee01}. 
\citet{vdm08} recently presented an estimate of M31's transverse 
velocity based on indirect arguments, using the known kinematics of 
satellite galaxies of M31 and the Local Group. Their estimate 
implies that $V_{\rm tan} \leq 56$ km s$^{-1}$, at 68.3\% confidence. 
In other words, M31 may be moving directly (radially) towards the 
Milky Way. This might have drastic implications for the future of 
the Milky Way, and even for the Sun itself \citep{cox08}. However, 
all theoretical and indirect estimates of $V_{\rm tan}$ make 
assumptions about the structure and equilibrium of the Local Group 
that may not hold true in our hierarchically-evolving universe. 
Therefore, only a direct PM measurement will yield a robust 
determination of the M31 velocity vector. 

The {\it Hubble Space Telescope} (\hst) is an observing platform in
space with unparalleled astrometric capabilities. Whereas the 
{\it absolute} astrometric accuracy of \hst\ is limited by the 
accuracy of the Guide Star Catalog \citep[$\sim0.2$ arcsec for 
GSC2.3;][]{las08}, the current onboard imagers of \hst\ (e.g., ACS 
and WFC3) are capable of measuring {\it relative} positions of 
multiple sources in a field to better than 0.5 mas. Therefore, it is 
possible to measure very accurate absolute PMs of stars by measuring 
their displacement over time relative to stationary background 
reference source(s) in the same field.

Quasi-stellar objects (QSOs) have conventionally served as reference
sources in \hst\ PM studies, due to their star-like appearance and
luminous nature. Such a strategy has been used in several studies for
measuring absolute PMs of Milky Way satellite galaxies
\citep[e.g.,][]{pia02,pia03,pia05,pia06,pia07,kal06a,kal06b,pia08}.
However, the accuracies thus achieved are not sufficient at the
distance of M31. This is due in part to drawbacks of using QSOs. Not
only must they be spectroscopically identified in advance, but their
distribution on the sky is sparse. For the small field of view of
\hst\ imagers, these limits require the observer to find a QSO behind 
the field of interest. If one exists, one must deliberately image the
field containing the QSO at two or more different epochs to measure
the displacement of the stars relative to the QSO. Even then, the PM
uncertainty is limited by the positional accuracy of a single QSO.

Instead of using QSOs, one can use background galaxies as reference
objects in \hst\ studies of absolute PMs. At the spatial resolutions
of interest here, these sources are also stationary due to their 
large distances. One great advantage of background galaxies is their
ubiquity: background galaxies are found in nearly every deep
astronomical image. Moreover, using many background galaxies can
provide higher accuracy than using a single QSO. While the individual
positional accuracy is somewhat poorer for galaxies, their much larger
number provides an important $\sqrt{N}$ advantage in averaging. For
example, \citet{mil06} and \citet{kal07} used background galaxies
to derive an accurate PM of the Galactic globular cluster NGC 6397, 
at the distance of $\sim 3$ kpc from the Sun. Similarly, \citet{bro10}
used background galaxies to determine the absolute PM of a 
hypervelocity star in the Milky Way halo.

Unlike the situation with QSOs, galaxy positions are difficult to
measure. One can adopt a simple centroiding algorithm, but such
algorithms have limited accuracy in the undersampled images of \hst's
wide-field detectors \citep{and00}. Furthermore, the resolved galaxies
and the unresolved stars would be expected to have different centroid
biases, leading to spurious apparent PMs. As a better alternative,
\citet{mah08} developed a template-fitting method to measure accurate
positions of resolved background galaxies. In this method, a 
tailor-made template is constructed for each individual object, and 
the same template is used to measure a consistent position for that 
object in each exposure. This is conceptually similar to the 
effective point spread function (ePSF) approach widely used for stars 
\citep{and00}. Even for bright background galaxies, the 
template-fitting method is able to measure positions at least twice 
as well as with simple centroids \citep{mah08}.

In this paper we present the first proper motion measurements for the
galaxy M31, based on new \hst\ ACS/WFC and the WFC3/UVIS imaging of
three fields with deep pre-existing data. We use the template-fitting
method of \citet{mah08} as our basis, but make several innovations in
the analysis to minimize systematic errors and ensure robustness of
the results. The final accuracy of only $12\ {\mu}$as yr$^{-1}$ is
significantly better than what has been achieved by \hst\ for other
nearby galaxies, and is comparable to what has been achieved for 
other Local Group galaxies using VLBA observations of water masers.

This paper is organized as follows. Section 2 describes the
observations. Section 3 describes the analysis steps that lead to the
PM measurements. Section 4 presents and discusses the PM results thus
obtained for the three target fields. Concluding remarks are
presented in Section 5. 

This is the first paper in a series of three. In Paper~II 
\citep{vdm12} we derive the velocity vector of the M31 center of mass 
by correcting the PM results presented here for the internal motions 
of stars within M31 and for the reflex solar motion. We then combine 
the result with other estimates to determine the Galactocentric 
velocity and orbit of M31, and use this to estimate the mass of the 
Local Group. In Paper~III (van der Marel et al. 2012b, in preparation) 
we use the results to study the future dynamical evolution of the 
Local Group, and the expected merging of the Milky Way-M31-M33 System.

\section{Data}\label{sec:Data}

\subsection{First-Epoch Data}
\label{sec:Data1}

\begin{figure}
\epsscale{1.0}
\plotone{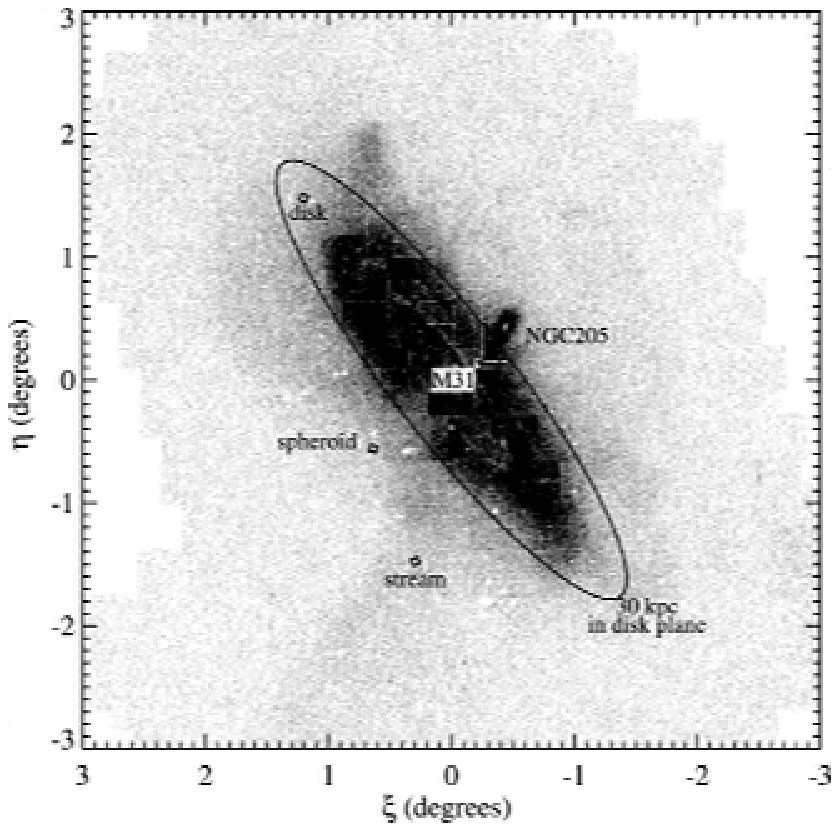}
\caption{Figure taken from \citet{bro06}. Appropriately scaled and 
         rotated boxes denote our three target fields (labeled). 
         The underlying gray shading represents the density of stars 
         from the map of \citet{fer02}. The ellipse marks the area 
         within 30 kpc of the galactic center in the inclined disk 
         plane (labeled). 
         \label{fig:TargetFields}}
\end{figure}

The data set we used for measuring the PM of M31 consists of \hst\
images of three fields obtained in two separate epochs. The fields 
are the {\bf SPHEROID}, {\bf OUTERDISK}, and {\bf TIDALSTREAM} fields 
of \citet{bro06} obtained under the science programs GO-9453 and 
GO-10265 (PI: T. Brown).  Figure~\ref{fig:TargetFields} shows the 
location of the three M31 target fields. The observation details for 
the first epoch data are given in \citet{bro06}; field coordinates 
and total exposure times are listed in their Table~\ref{tab:ObsLog}. 
In brief, the target fields were observed between December 2002 and 
January 2005 in two filters (F606W and F814W) with HST ACS/WFC to 
create color-magnitude diagrams (CMDs) that reach well below the M31 
main sequence turn-off. For each field, all exposures were obtained 
within 40 days so they can be safely treated as a single epoch of 
data. Total per-field per-filter exposure times ranged from 53-161 
ksec. Individual images had exposure times ranging from 1,100 to 
1,300 sec, and were dithered such that no two exposures placed a 
star on the same pixel. In addition to the whole-pixel offsets, the 
dithering strategy also included sub-pixel shifts so that every 
object was observed at a range of locations relative to the pixel 
boundaries; such ``pixel phase'' coverage is critical for our 
program. The very deep first-epoch F606W data were used by us for 
constructing super-sampled stacked images, identifying stars and 
galaxies, building templates for the template-fitting method, and 
providing reference positions of stars and galaxies, as described 
in Section~\ref{sec:MeasuringPM} below.

\subsection{Second-Epoch Data}
\label{sec:Data2}

Our second-epoch data were observed between January 2010 and August
2010 using ACS/WFC and WFC3/UVIS in F606W (HST program GO-11684, PI:
van der Marel). We designed our observations to minimize random and
systematic errors with the following strategies in mind. 
Pre-analysis of the first-epoch data revealed that the F606W filter 
provides slightly better astrometric handle on the background 
galaxies than the F814W filter, so we chose to obtain second-epoch 
images only with F606W. Furthermore, we decided to observe the three 
target fields with two different instruments instead of just one.
This yields comparable random error, but allows additional 
consistency checks on our PM results. Since the purpose of the 
second-epoch observations was astrometry, and not deep photometry, 
the total second-epoch exposure time was much less than in the 
first-epoch. We observed each target field for two orbits (four
half-orbit exposures) with WFC3/UVIS and one orbit (two half-orbit
exposures) with ACS/WFC. Individual images had exposure times 
ranging from $\sim$1,300 to 1,500 sec, slightly longer than in the 
first epoch, and were dithered in similar fashion. The baselines 
between the two epochs are in the range of 5--7.5 years.

For the ACS observations, we used the same orientations and
coordinates as used by the first-epoch observations to maximize the
overlapping area and so that any errors in the static distortion
solution would naturally cancel out. Due to different sets of guide
stars being used between the first- and second-epoch observations, 
the second-epoch ACS images were slightly offset and rotated with 
respect to the first-epoch images. However, the offsets are all 
within a negligible fraction of the ACS/WFC field of view (FOV), and 
the orientations are within 0\fdg1 of each other.

For the WFC3 observations, we pointed the telescope so that the
resulting images roughly share the same centers with the first epoch,
but the WFC3 images were rotated by roughly 45\degr\ with respect to
the ACS images.  This was to place the parallel ACS/WFC fields
overlapping with the first epoch parallel WFPC2 images, but the
parallel fields will not be discussed in this paper as they are not 
useful for astrometry at the accuracy that we need. Because the FOV
of WFC3/UVIS is 64\% of that of the ACS/WFC, the observed WFC3 images
are fully contained within the ACS images despite the $\sim45$\degr\
difference in their orientations. The details of our second-epoch 
data are summarized in Table~\ref{tab:ObsLog}.

\section{Analysis}
\label{sec:MeasuringPM}

\subsection{Overview}
\label{sec:Overview}

Measurement of proper motions involves measuring the displacement
between the position of an object at one time and its position at
another. If this displacement can be measured with respect to objects
for which we know the absolute motion, then we can obtain absolute
PMs. In our case, we will measure the displacement of M31 stars with
respect to the background galaxies in the field to obtain an absolute
PM. Our strategy is to first align the stars in the first- and 
second-epoch images, restricting the alignment to those stars 
confirmed to belong to M31. Then we measure the average displacement 
of the background galaxies with respect to this moving frame of 
reference. The negative of this relative displacement yields the mean 
absolute PM of the M31 stars. This measurement of the mean does not
require knowledge of the exact PM {\it distribution} of the M31 stars. 
This distribution can be complex, because different structural 
components of M31 contribute to each field. However, to transform
the mean PM of the M31 stars in a field to an estimate of the M31 
center of mass, one does need to construct a model for the internal 
kinematics of M31, which is addressed in Paper~II of this series.

The overall PM derivation process is summarized below.
\begin{enumerate}

\item We first create a high-resolution stacked image using the deep 
and well-dithered first-epoch images for each target field
using the distortion-corrected first exposure as the reference frame.

\item Stars and galaxies are then identified from the stacked image 
for each field, and photometric measures and the CMD are used to 
specifically identify M31 reference stars.

\item We construct a template for each star and galaxy from the 
high-resolution image. 

\item We then use the template to measure in a consistent fashion a 
position for each star/galaxy in each individual exposure
of each epoch.

\item We redefine the first-epoch reference frame using the 
template-based positions of the stars, and determine the average 
first-epoch position of each galaxy in this frame. 

\item We determine the template-measured positions of the stars in
each of the six second-epoch exposures for each field and use them to
transform the template-measured positions of the galaxies into the
first-epoch frame. 

\item We then take the difference between the second- and 
first-epoch positions of the galaxies to obtain the relative 
displacement of the galaxies with respect to the M31 stars.

\item Finally, we multiply the relative displacements of the galaxies 
by $-1$ to obtain the mean absolute displacement of the M31 stars, 
since in reality the galaxies are stationary and the stars are moving.
Division by the time baseline turns the displacements into PMs.

\end{enumerate}

\noindent The following sections provide the details of each step.

\subsection{Initial Processing of Individual Exposures}
\label{sec:InitialProcessing}

We downloaded all the required data in the form of {\tt \_flt.fits}
images from the STScI \hst\ archive. These images are already
bias-subtracted, dark-subtracted, and flat-fielded by the STScI
data-reduction pipelines using the best available calibration data at
the time of data inquiry.  Each multi-extension {\tt \_flt.fits} file
was converted to a single 4,096$\times$4,096 image where the top chip
scene is directly abutted to the bottom chip scene. This single-format
image allows a more efficient way of handling the data, and all
computer software we used throughout our analysis were specifically
programmed to deal with this image format.

The \hst\ detectors suffer from significant degradation in 
charge-transfer efficiency (CTE) as time progresses. This impacts the 
astrometry of objects by effectively shifting the centroid position 
of the flux distribution. This can be a particularly serious problem 
for our second-epoch ACS/WFC data, taken almost 9 years after the 
installation of ACS. A pixel-based CTE correction routine has 
recently been developed by \citet{and10} for the ACS/WFC, and 
\citet{bro10} demonstrated that this correction works very well for 
astrometric purposes. All first-epoch ACS/WFC images were processed 
through this routine. For the second-epoch ACS/WFC images, we used a 
modified version of the CTE correction code that also included a 
correction for $X$-CTE, in addition to the publicly available 
correction for $Y$-CTE. We found that the $X$-CTE correction made 
very little difference to the positions measured in the images, 
compared to the other uncertainties in the results.

A CTE correction routine was not available for the WFC3/UVIS at the
time of writing this paper, so no correction was made to the WFC3
images. However, since at the time of the observations, UVIS had 
spent only $\sim10$\%\ as much time as ACS in the space environment, 
the CTE degradation of our UVIS images is expected to be quite small.
Although UVIS appears to be losing charge-transfer efficiency faster 
than ACS did, the studies that show this have focused on 
low-background images \citep[e.g.,][]{bag11}. As our backgrounds are 
quite high, we expect CTE losses to be very low for our images.
Hence, the CTE degradation should only have a minimal effect on our 
WFC3 PM measurements. A direct post-facto consistency check of this 
is provided by comparison our ACS/WFC and WFC3/UVIS PMs derived for 
the same fields (see Section~\ref{sec:checks} below).

As the final step of the initial processing of individual images, we
measured each star using the {\tt img2xym\_WFC\_9x10} program
\citep{and06} on the ACS/WFC images, and using a similar program on
the WFC3/UVIS images.  Both programs utilize library PSFs to 
determine a position and a flux for each star. We will use these 
PSF-based star positions to provide our initial handle on the 
coordinate transformations from one image to another.

\subsection{Distortion Corrections}
\label{sec:DistCor}

The positions of objects on the \hst\ detectors must be corrected 
for geometric distortions before they can be used for any astrometric
measurement. For the ACS/WFC detector, the distortion-correction
solutions by \citet{and05} have been used in several astrometric
studies. These solutions are known to be better than 0.01 pixel or
$\sim$0.5 mas, and were shown to be stable between 2002 and 2007
\citep{and07,vdm07}. These available solutions were directly applied
to the positions measured in our first-epoch ACS/WFC images.

While reducing data for another program (GO-11677, PI: H. Richer), 
one of us (J.A.) found that the ACS distortion solution appears to 
have changed by $\sim0.005$ pixel relative to the solution before the 
{\it HST Service Mission 4} (SM4). This may be related to a similar
variation of the PSF (Anderson et al., in prep.). It is unclear what
caused these variations, but they appear to be stable over time. We
used the GO-11677 data set to develop a correction to the pre-SM4
ACS/WFC distortion solution and applied it to the second-epoch data. 
For the WFC3/UVIS data, we used the distortion corrections provided 
by \citet{bel11}. These corrections are better than 0.008 pixel or 
$\sim$0.3 mas and appear to be stable over the UVIS lifetime so far.

The accuracies of the distortion corrections are comparable to our 
final measured motions (as will be shown later). However, geometric
distortion affects both star and galaxy positions similarly, and
hence drops out in a differential measurement to lowest order. This 
is true in particular if each galaxy position is measured with
respect to that of nearby stars that fall on the same area of the
detector. This is what we do in Section~\ref{sec:DerivingPMs} below, 
to ensure that higher order geometric distortion correction residuals
do not affect our final results.

\subsection{Stacked Images}
\label{sec:StackedImages}

The deep and well-dithered first-epoch ACS/WFC data are well-suited
for constructing high-resolution stacked images. To do this, we first
apply the distortion corrections (see Section~\ref{sec:DistCor}) to
the positions of stars measured in Section~\ref{sec:InitialProcessing}. 
We then adopt the first exposure of each field as the frame of 
reference, cross-identify stars in this exposure and stars in the 
other exposures, and use their distortion-corrected positions to 
construct a six-parameter least-squares linear transformation between 
the two frames. The six parameters involve x-y translation, scale, 
rotation, and two components of skew. The need for these linear 
transformations is discussed in Section~3.6.4 of \citet{andvdm10}. 
We use these transformations to convert the star positions measured 
in the various images into the reference frame, giving us many 
estimates of the position of each star in the reference frame. 
We average these positions to refine the reference-frame positions 
for all the stars, and use these new average positions to improve the 
transformations (which had initially been based on the positions as 
measured in the first frame itself). This procedure is iterated 2--3 
times to improve the positions of both the stars and the 
transformations.

\begin{figure}
\epsscale{1.0}
\plotone{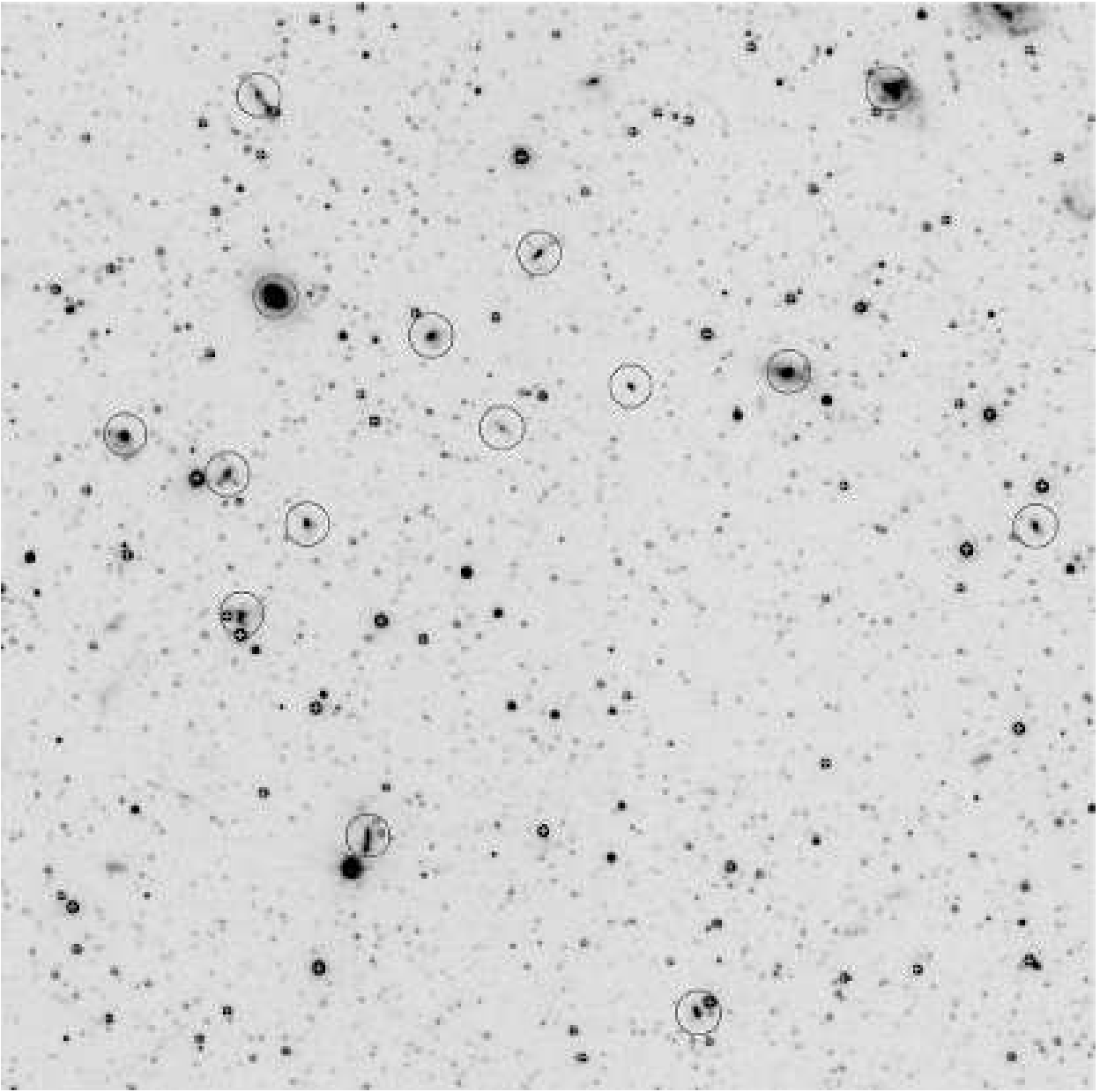}
\caption{A 25\arcsec$\times$25\arcsec\ portion (1.5\% of the total 
         image area) of the stacked image for the {\bf SPHEROID} 
         field. Background galaxies used as positional references are 
         enclosed by red circles, while M31 stars that pass the 
         selection criteria of Section~\ref{sec:StarIdentification} 
         are marked with green plus signs. 
         \label{fig:StackedImage}}
\end{figure}

We use the star-based transformations to construct the stacked images.
In order to get better sampling, we super-sample the stacked image by
a factor of 2 relative to the native ACS/WFC pixel scale. The
image-stacking process we used is similar to the commonly used {\tt
Drizzle} algorithm \citep{fru02} with a point kernel, except that we
included an iterative procedure to regularize the sampling in a 
manner similar to {\tt iDrizzle} \citep{fru11}. The procedure 
involves no deconvolution, and as such the resulting image at every 
point simply represents the flux that an actual {\tt \_flt} image 
pixel would receive if it were centered at that location in the frame. 
It is this property that allows us to construct empirical templates 
(see Section~\ref{sec:TemplateFittingMethod}) for our stars and
galaxies. Figure~\ref{fig:StackedImage} shows a
25\arcsec$\times$25\arcsec\ portion of the 2$\times$ super-sampled
stacked image for the {\it SPHEROID} field with M31 stars and
background galaxies identified, as described in the following
sections.

\subsection{Identifying Stars and Galaxies}
\label{sec:StarGalaxyID}

All of our target fields are dominated by M31 stars, but there are 
also other sources, such as foreground stars and background galaxies. 
Since our strategy for deriving the PM of M31 is to measure the 
displacement of the background galaxies relative to the co-moving 
frame of reference defined by the M31 stars, it is important to 
accurately identify both background galaxies and M31 stars before we 
proceed any further.

\subsubsection{Identifying M31 Reference Stars}
\label{sec:StarIdentification}

\begin{figure}
\epsscale{1.0}
\plotone{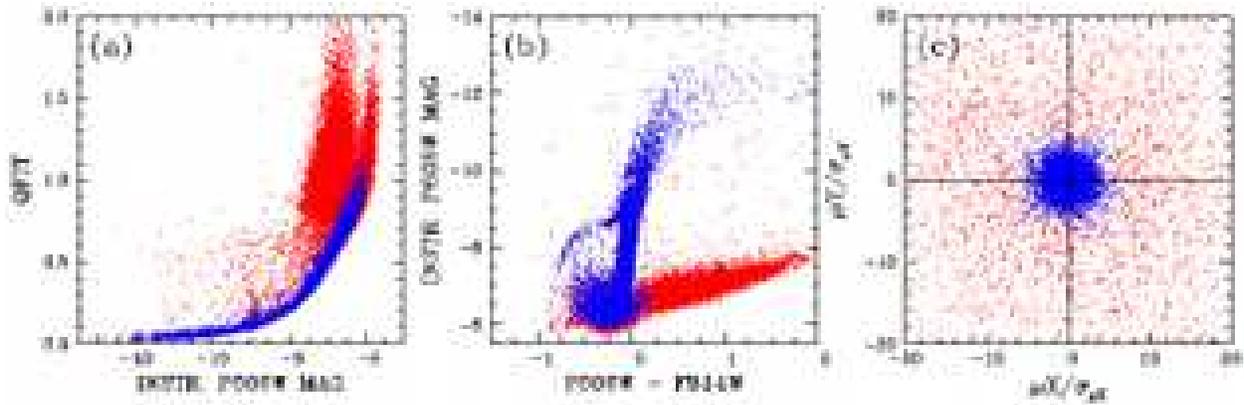}
\caption{Selection of M31 stars in the {\bf SPHEROID} field based on
         (a) the quality-of-fit ({\tt QFIT}) parameter,
         (b) the color-magnitude diagram, and
         (c) relative proper motion (divided by the proper motion 
         error for better scaling) with respect to the average proper 
         motion. In all three panels, blue points are stars that pass 
         all three cuts, while red points are objects that fail to 
         pass at least one cut. Most objects in red located at the 
         bright end of the color-magnitude diagram were rejected 
         because of their relative proper motions. These are likely
         foreground stars.
         \label{fig:StarID}}
\end{figure}

The selection of M31 stars was carried out following the procedure
below.  For each target field, we use the star list compiled in
Section~\ref {sec:StackedImages}. Our initial star lists include only
sources that are found independently in a large number of first-epoch
F606W exposures (typically, $> 35$\%); as such, the list is almost
entirely free of cosmic rays and image defects. There are, however, a
large number of resolved sources, which could be either blended stars
or extended background galaxies. To identify them, we made use of the
mean quality-of-PSF-fit parameter ({\tt QFIT}) reported by the {\tt
img2xym\_WFC\_9x10} program (see Figure~\ref{fig:StarID}a). Selecting 
the objects with low values of {\tt QFIT} allows us to filter out 
extended objects as well as stars that are too close to other stars 
to provide accurate position measurements. To ensure that all of the 
stars in the list are M31 members, we reduced the F814W images and
constructed a CMD (see Figure~\ref{fig:StarID}b). From this, we
selected stars that lie roughly on the M31 sequences. Although 
small in number, field stars in the Milky Way halo may be included 
in our selection, but we believe that most of them will be filtered 
out in our next step of selection. Finally, reference stars were 
selected based on their lack of motion with respect to the other M31
stars. This was done by aligning the second-epoch star positions with
the first-epoch positions and iteratively rejecting the objects that
have moved between the epochs (see Figure~\ref{fig:StarID}c). For any
given ACS/WFC target field, the M31 stars should all be moving 
towards the same direction in space. Differential motions due to the 
internal kinematics of M31 within a single ACS/WFC field should be 
negligible compared to the observational uncertainties (see Paper~II 
for details). We note that, in principle, better detection and 
photometry of fainter stars can be achieved by measuring stars 
directly from the stacked images as has been demonstrated by 
\citet{bro06}. However, as our main goal is doing astrometry, we are 
only interested in stars for which positions can be reliably measured 
in {\it individual exposures}. Our final lists of M31 reference stars 
include $\sim 10,000$ stars for the {\bf SPHEROID} and {\bf OUTERDISK} 
fields, and $\sim 5,000$ stars for the {\bf TIDALSTREAM} field.

\subsubsection{Identifying Background Galaxies}
\label{sec:GalaxyIdentification}

A quick visual inspection of the super-sampled stacked images of our
target fields reveals that there are hundreds of background galaxies. 
We have already identified sources using the {\tt img2xym\_WFC\_9x10} 
program, but because that program is specifically designed to measure 
stars, it neglects many of the extended sources. For this reason, 
{\tt SExtractor} \citep{ber96} was separately run on the stacked 
images to detect and measure extended sources. We generated a 
candidate list of galaxies by selecting sources from the 
{\tt SExtractor} output mainly based on their {\tt MAG\_AUTO}, 
{\tt CLASS\_STAR}, and {\tt FLUX\_MAX} parameters. For each field, 
this candidate list included more than 1,000 sources, but we found 
that many of the candidates are in fact multiple stars clustered 
together. We carefully identified sources in the candidate list one 
by one, basing our judgment on their 2-d contours and 3-d surface 
profiles. Whenever the identification was unclear, we excluded the 
source from our selection to stay on the conservative side. We note 
that we desire the background galaxy list to be as free of 
contamination by stars as possible because the PM measurements are
sensitive to the reference sources we choose to use. Finally, we
considered only objects for which the template-fitting method,
described in the next section, yields position uncertainties of less
than 0.25 pixel in both $X$ and $Y$. The final lists contain 
368, 339, and 374 galaxies for the {\bf SPHEROID}, {\bf OUTERDISK}, 
and {\bf TIDALSTREAM} fields, respectively.

\subsection{Measuring Positions of Objects with Templates}

Our goal is to measure accurate PMs. The most crucial part of our
analysis is therefore to measure the positions of objects in the
individual images as accurately as possible. Simple methods such as
fitting the objects with a two-dimensional Gaussian function or
finding the flux centroid are accurate only if the objects have
well-sampled peaks. This is not the case for most of the background
galaxies in our target fields. To measure the positions of objects, 
we therefore adopt the template-fitting method developed by
\citet{mah08}. 

\subsubsection{The Template-Fitting Method}
\label{sec:TemplateFittingMethod}

The details and general diagnostics of the template-fitting method 
are documented in \citet{mah08}. Here we summarize only the basic 
concepts behind the method. The distortion corrections in
Section~\ref{sec:DistCor} and the linear transformations derived in
Section~\ref{sec:StackedImages} allow us to associate a position
$(x_{r}, y_{r})_{j}$ in a given individual image $j$ with a position
$(x_{m}, y_{m})$ in the stacked image, and {\it vice versa}. We are
thus able to build a model of what the galaxy or star should look 
like in an individual image by proper sampling of the stacked image.

The goal is to measure in consistent fashion a position for each
object (star or galaxy) in each individual exposure. To do this, we
construct a template model for each object, and use it to measure 
that object in every exposure. The location within the template that 
we will define to be its center is arbitrary, since in the end we 
care only about differences in position. So we define the center of 
the brightest pixel in the stacked image to be the object center (or
``handle'' in the parlance of Mahmud \& Anderson). With the center so
defined, we interpolate the stacked image at a super-sampled array of
points, with the center at (0,0) but with a pixel orientation and
spacing that corresponds to the transformed exposure's {\tt \_flt}
coordinate system. The result is a template that can be interpolated
at an array of locations to tell us how much flux we would expect in
an array of {\tt \_flt} pixels for a presumed object center. The
template is custom-made for each individual exposure, based on the
mapping of that exposure's pixel coordinate system into the master
frame. The template centers are all the same.

For each object in each exposure, we use the previously established
transformations to identify the location of the object in the 
exposure to within one pixel. We then evaluate an array of trial 
locations for the center, covering an area of 1$\times$1 pixels with 
a spacing of 0.01 pixel. At each trial location, interpolation of the 
template with a bicubic interpolation algorithm tells us how much 
flux we should expect to measure in each pixel of the exposure. 
We compare this model to the actual observed pixel values to obtain 
a quality-of-fit quantity,
\begin{eqnarray}
      qfit_{ij} = {{\Sigma_{\rm pixels} \> | OBS - MODEL_{ij} | } 
                    \over {\Sigma_{\rm pixels} \> OBS}},
\label{eq:qfit}
\end{eqnarray}
for each trial offset $(i,j)$. The position of the galaxy or star 
then corresponds to the location of the template center that gives 
the minimum $qfit_{ij}$.

In Section~\ref{sec:InitialProcessing} we used empirical ``library''
PSF fits to the stars to define the transformations for construction
of the stacked images. However, for the final PM measurements we use
the template procedure for both galaxies and stars. This minimizes 
any potential for systematic errors due to differences in measurement
techniques between different types of objects. The sums in
equation~(\ref{eq:qfit}) were chosen to extend over the $5\times5$
pixel raster centered on the object's brightest pixel in the
individual exposure. The size of this region was deliberately chosen 
to be small, since most of the positional information for the stars 
and compact galaxies of interest here is contained in the central 
core pixels.

\subsubsection{First-Epoch ACS Data}
\label{sec:FirstEpochACS}

\begin{figure}
\epsscale{1.0}
\plotone{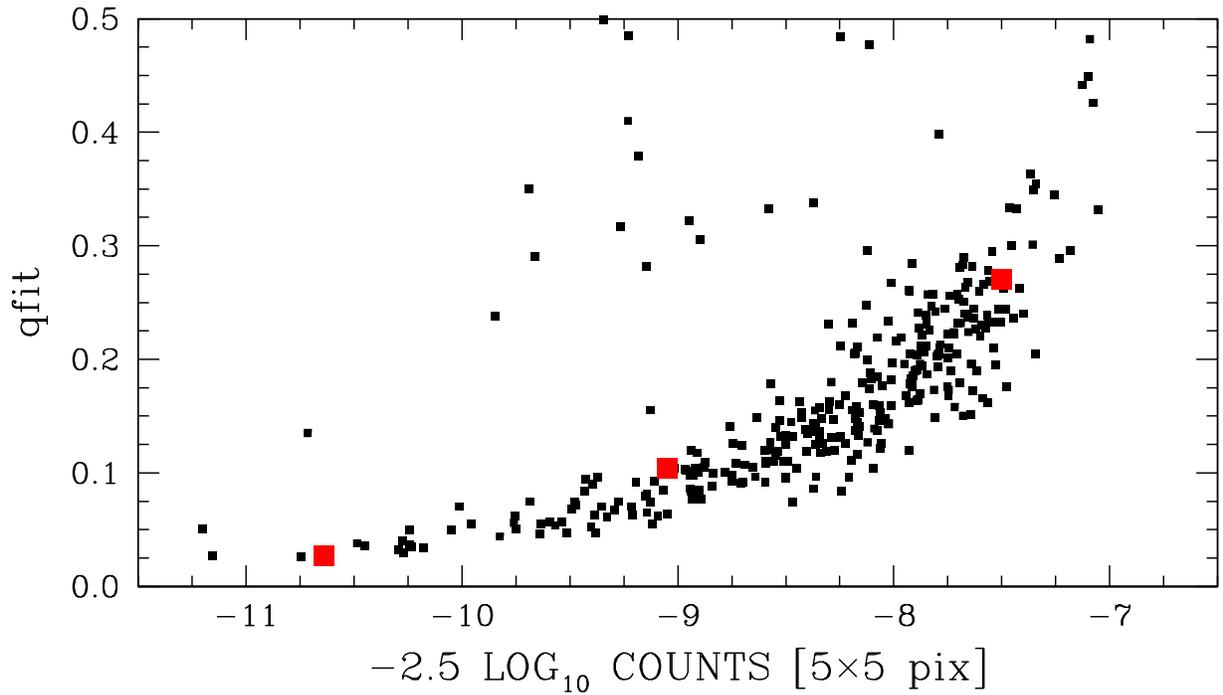}
\caption{Best fit $qfit$ (as defined in
         Section~\ref{sec:TemplateFittingMethod}) versus instrumental 
         magnitude for the background galaxies in one of the 
         first-epoch images ({\tt j8f801caq}). The quantity $qfit$ is 
         a measure of the typical flux residual of the template-fit 
         normalized by the total flux. It therefore behaves as 
         $(S/N)^{-1}$, so that brighter objects yield smaller values 
         of $qfit$. The red squares mark the objects shown in 
         Figure~\ref{fig:TemplateFits}.
         \label{fig:TemplateFittingACSEP1}}
\end{figure}

\begin{figure}
\epsscale{1.0}
\plotone{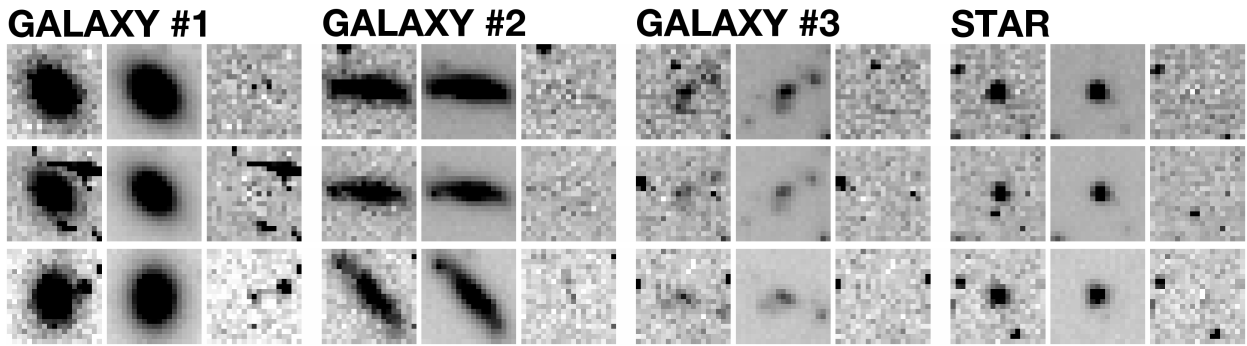}
\caption{Each row shows example results of the template-fitting
         method. Top row: a first-epoch ACS/WFC exposure; middle row: 
         a second-epoch ACS/WFC exposure; bottom row: a second-epoch 
         WFC3/UVIS exposure. Each row shows results for four 
         different objects. The first three objects (denoted GALAXY 
         \#1, \#2, and \#3), viewed from left to right, are galaxies 
         of decreasing brightness. The rightmost object 
         (denoted STAR) is a star with similar brightness to the 
         second galaxy. Note that the bottom row images are rotated 
         45\degr\ clockwise with respect to those in the top and 
         middle rows because of the difference in orientations 
         between ACS and WFC3 observations (see 
         Section~\ref{sec:Data2}). For each object in each image we 
         show three images: the observed pixels; the best-fit 
         template; and the residual of the two. The images show a 
         $19\times19$ pixels patch for illustration, even though the 
         template-fitting was done using only the inner $5\times5$ 
         pixels. The three galaxies in the top row are the same ones 
         for which the $qfit$ value is indicated with a red square in 
         Figure~\ref{fig:TemplateFittingACSEP1}.
\label{fig:TemplateFits}}
\end{figure}

We applied the template-fitting method to all stars and galaxies in
the first-epoch ACS/WFC data, to determine their positions in each
exposure. To illustrate the quality of the template-fitting method, 
we show in Figure~\ref{fig:TemplateFittingACSEP1} the best fit $qfit$ 
(as defined in Section~\ref{sec:TemplateFittingMethod}) versus
instrumental magnitude for the background galaxies in one of the
first-epoch images. The top row of Figure~\ref{fig:TemplateFits} 
shows the data, template fits, and residuals for three galaxies of 
different brightness, and for one star.

\subsubsection{Second-Epoch ACS Data}
\label{sec:SecondEpochACS}

The stacked images constructed in Section~\ref{sec:StackedImages}
represent the astronomical scene convolved with the average effective 
PSF {\it appropriate for the first-epoch data}. The effective PSF 
represents the convolution of the instrumental PSF with the 
sensitivity profile of a pixel. The astronomical scene for any given 
object should not change between our epochs of data, but the 
effective PSF might change. This is certainly the case for the 
second-epoch WFC3/UVIS data, discussed below, since both the 
instrumental PSF and the pixelization are different than for the 
first-epoch ACS/WFC data. However, we also found that the effective 
PSF for the second-epoch ACS/WFC data was not the same as for 
first-epoch ACS/WFC data. Even though the pixelization remained the 
same, we found that a subtle change occurred in the ACS PSF since 
SM4 (see Section~\ref{sec:DistCor}).

It is important that any change in the effective PSF be explicitly
accounted for in the analysis. PSF changes can introduce centroid
shifts, and these shifts can be different for point sources and for 
extended sources. Our technique for measuring the M31 PM relies on 
positional differences between stars and background galaxies. 
PSF changes can therefore produce spurious PM results.

In order to deal with the differences in the effective PSF between
epochs, we model each object in a given second-epoch image as follows.
First, we select bright and isolated stars from the M31 reference 
star list compiled in Section~\ref{sec:StarIdentification} 
, and use them to derive a 7$\times$7-pixel kernel that accounts for 
differences in effective PSF between the stacked image and the 
individual exposure. Next, we interpolate the first-epoch template 
onto the {\tt \_flt} pixel grid of the second epoch image, as we did 
in Section~\ref{sec:FirstEpochACS}. We then convolve the template 
with the kernel derived in the earlier stage, and finally evaluate 
the goodness-of-fit quantity $qfit$ to find the object position that 
provides the minimum $qfit$. The middle row of 
Figure~\ref{fig:TemplateFits} shows the data, template fits, and
residuals for three galaxies of different brightness, and for one
star, obtained with this procedure for one of the second-epoch 
ACS/WFC images ({\tt jb4404vuq}).

The complexity in this procedure lies in the determination of the
kernel. Each kernel has 49 free parameters. However, it is 
constrained by 5$\times$5-pixel patches centered around {\it all} the 
objects in each image. This is an over-determined problem that can be 
solved to find the best-fitting kernel. We cast the problem into the 
form of a least-squares matrix equation, and solved for the kernel 
using the observed pixel values for several thousand bright and 
isolated stars distributed throughout the target fields. The solution 
was obtained through {\it singular value decomposition} \citep{pre92}. 
To fit the kernels, we first determined the stellar positions of the 
bright and isolated stars without any kernel. Then we kept those 
positions fixed, and optimized the kernels.

We ended up using a single kernel for each second-epoch image. 
We experimented with using different kernels for different parts of 
each image (to deal with potential field-dependent PSF variations), 
but found that this did not change the main results. In deriving the 
kernel, we recognized that there is a degeneracy between shifting the 
kernel and shifting the astronomical scene. We removed this 
degeneracy by constructing the pixel residuals that went into the 
kernel with respect to the library-PSF-measured position (see 
Section~\ref{sec:InitialProcessing}) for each star. This ensured that 
the kernel would reproduce on average the PSF-based positions for the 
stars, and that it would properly account for changes in source shape 
due to known PSF variations between epochs. Either way, the 
degeneracy does not affect absolute PMs that are based on relative 
differences in position between stars and background galaxies.

\begin{figure}
\epsscale{1.0}
\plotone{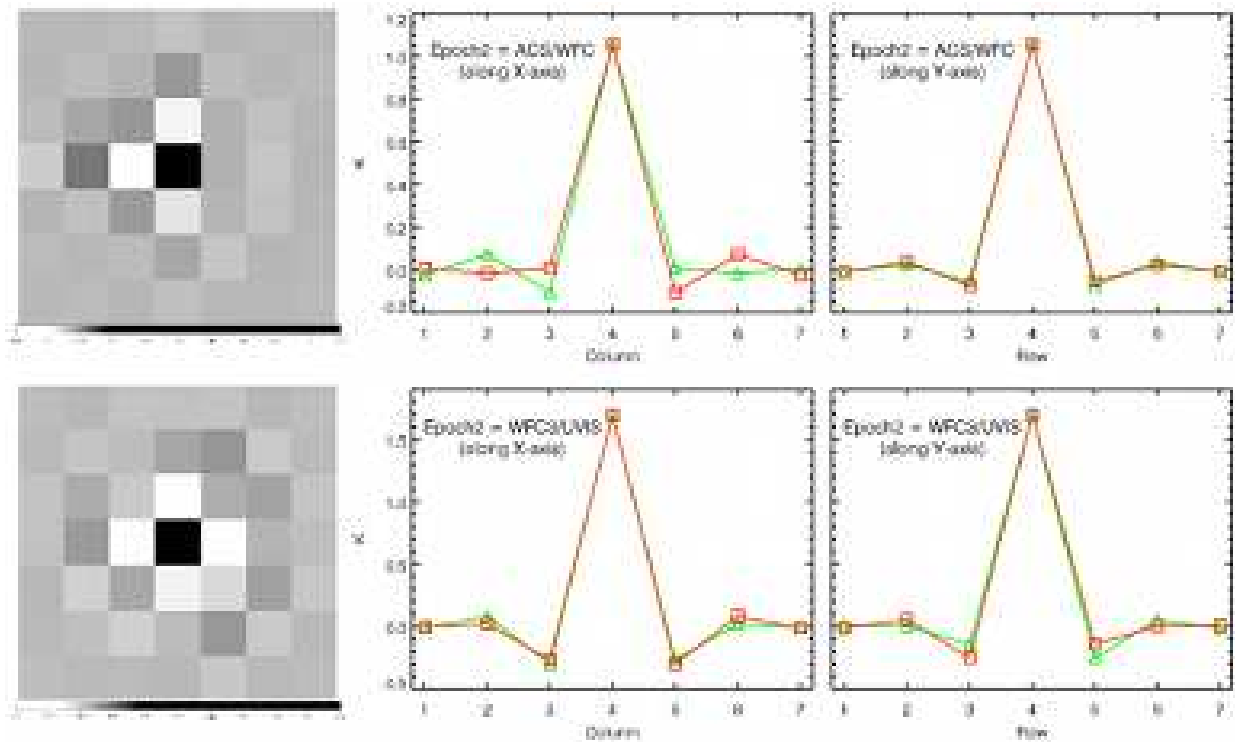}
\caption{Grayscale images (left panels) and one-dimensional cuts 
         (mid and right panels) of the convolution kernels derived 
         for the second-epoch ACS/WFC (top panels) and WFC3/UVIS 
         (bottom panels) data of the {\bf SPHEROID} field. The 
         one-dimensional cuts show in green the central row (mid 
         panels) and the central column (right panels) of the kernel. 
         Red squares show the same values in reverse order 
         (equivalent to flipping the plots about the center along the 
         abscissa). Comparing the plots in different symbols (and 
         colors) provides a measure of asymmetries in the kernels, 
         which could affect astrometry if left uncorrected.
\label{fig:kernel_diag}}
\end{figure}

The top panels of Figure~\ref{fig:kernel_diag} show the average of 
the derived normalized kernels for the second-epoch ACS/WFC images. 
Slight asymmetries are evident, primarily in the $X$ direction. 
Without our kernel-based corrections, such asymmetric PSF differences 
would bias astrometry.

\subsubsection{Second-Epoch WFC3 Data}
\label{sec:SecondEpochWFC3}

Our second epoch WFC3/UVIS images are rotated by $\sim45$ deg with
respect to the ACS images. Because we have many bright M31 stars in
the field, we are able to derive six-parameter linear transformations
that relate the undistorted WFC3 positions to the stacked image
without difficulty.

To determine the positions of all objects in the field we used the
same kernel-based template-matching approach as for the second-epoch
ACS data. In this case, the kernel accounts not only for PSF
variations, but also for variations in the PSF size and orientation
between epochs. The bottom row of Figure~\ref{fig:TemplateFits} shows
the data, template fits, and residuals for three galaxies of 
different brightness, and for one star, obtained with this procedure 
for one of the second epoch WFC3/UVIS images ({\tt ib4401s1q}).

The kernel size for the WFC3/UVIS data was chosen to be the same as
for the ACS data, and we used over 1,000 stars to derive the kernels. 
The bottom panels of Figure~\ref{fig:kernel_diag} show the average of 
the derived normalized kernels for the second-epoch WFC3/UVIS images. 
There are slight asymmetries in both coordinate directions. This 
differs from the situation for the second-epoch ACS/WFC data, because 
the WFC3/UVIS detector is rotated by $\sim 45^{\circ}$ in the HST 
focal plane with respect to the ACS/WFC detector. Also, the symmetric 
part of the kernels differs between the second-epoch ACS/WFC images 
and the second-epoch WFC3/UVIS images. This is because of the 
difference in PSF shapes between the different detectors. Our 
kernel-based approach minimizes potential systematics arising from 
measuring positions of stars and galaxies in images with different 
PSF characteristics.

Our PM measurement technique for each field compares the
high-resolution stacked image from the first-epoch to the individual
exposures from the second epoch. The derived kernels account for
variations in the average PSF between the epochs, as well as
exposure-to-exposure PSF variations within the second epoch. We do
not compare second-epoch exposures to individual first-epoch 
exposures, but only to the high-resolution stacked image. The stacked
image is constructed from the first-epoch images, so on average, the
PSF does not change between the stacked image and the individual
first-epoch images. Therefore, any exposure-to-exposure PSF variations
that may exist within the first epoch do not affect the derived PMs. 
For this reason, there was no need to study or correct explicitly PSF 
variations within the first epoch data set.

\subsection{Reference-Frame Positions and Positional Uncertainties}
\label{sec:RefPos}

The processes outlined above lead to accurate positions of stars and
galaxies in each individual exposure. We align the star positions 
between exposures using the same iterative approach described in
Section~\ref{sec:StackedImages}, but this time using the newly 
derived positions based on the template-fitting method instead of the 
initial library-PSF-based positions. For each exposure for a given 
field, this procedure yields a six-parameter linear-transformation 
with respect to the first exposure in the first epoch for that field. 
For all objects (stars and background galaxies) in all exposures, we 
apply the known geometric distortion solutions and these 
linear-transformations, to obtain the position in the 
distortion-corrected frame of the first exposure (i.e., 
{\tt j8f801abq}, {\tt j92c01b4q}, and {\tt j92c28ccq} for 
{\bf SPHEROID}, {\bf OUTERDISK}, and {\bf TIDALSTREAM}, respectively). 
These distortion-corrected frames serve as our reference frames.

\begin{figure}
\epsscale{1.0}
\plotone{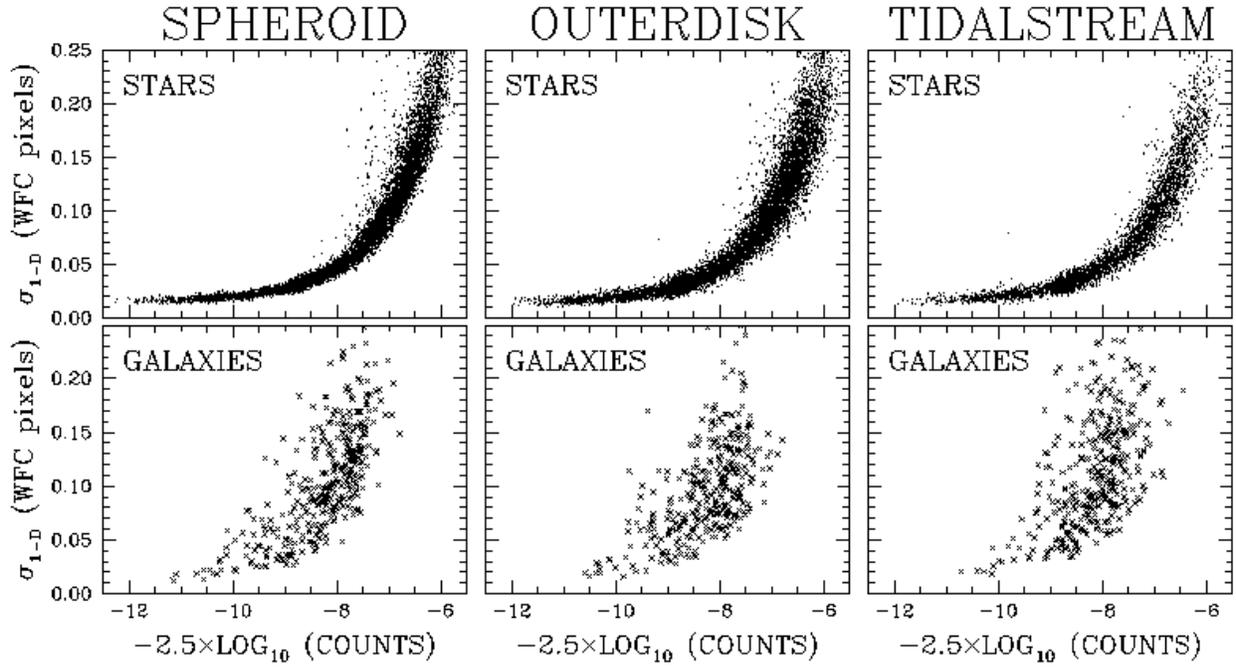}
\caption{The total one-dimensional positional error per exposure as a
         function of instrumental magnitude, for M31 stars 
         ({\it upper panels}) and background galaxies ({\it lower 
         panels}). The error is defined as $\sigma_{1-D} = 
         \sqrt{{1 \over 2}({\sigma_{X}}^{2} + {\sigma_{Y}}^{2})}$. 
         Here $\sigma_{X}$ and $\sigma_{Y}$ are the per-coordinate 
         RMS residuals with respect to the average, for the multiple 
         first-epoch measurements.
\label{fig:plot_sigxy}}
\end{figure}

In the first-epoch data we have many exposures. This yields multiple
determinations for the position of each star or galaxy in the
reference frame. The RMS scatter among these determinations provides
a measure of the random positional uncertainty in a single 
measurement for each object. Figure~\ref{fig:plot_sigxy} shows the 
one-dimensional errors in positions for M31 stars (upper panels) and 
background galaxies (lower panels) as a function of instrumental 
magnitude (defined as $m_{\rm instr} = -2.5 \log[{\rm electrons}]$) 
in the three separate target fields. Brighter objects have a higher 
signal-to-noise ratio than fainter objects, and therefore have more
accurately determined positions. Stars are more compact than 
background galaxies, and therefore also have more accurately
determined positions. At fixed magnitude, the scatter in positional 
accuracy is larger for the galaxies than for the stars. This is 
because galaxies have a large variation in sizes and shapes, with the 
morphology also depending on wavelength. The dependence of 
astrometric accuracy on size and shape was discussed in 
\citet{soh10}, and the dependence on wavelength was discussed in 
\citet{mah08}. At the bright end in Figure~\ref{fig:plot_sigxy}, the 
per-exposure accuracies level off at 0.01--0.02 WFC pixels for both 
stars and galaxies. When many measurements are averaged, the 
positional errors decrease as $1/\sqrt{N}$.

\subsection{Proper Motions of Individual Objects}
\label{sec:DerivingPMs}

Using the above procedure,we determine for each object its average 
first-epoch position in the reference frame. In a similar way, we 
measure template-based positions for each second-epoch exposure, 
correct these positions for distortion, and transform them into the 
reference frame using the template-measured star positions. We then 
compare these positions with the first-epoch averages to estimate the 
PM displacement for that object. Division by the time baseline of the 
observations then yields an estimate of the actual PM.

\begin{figure}
\epsscale{1.0}
\plotone{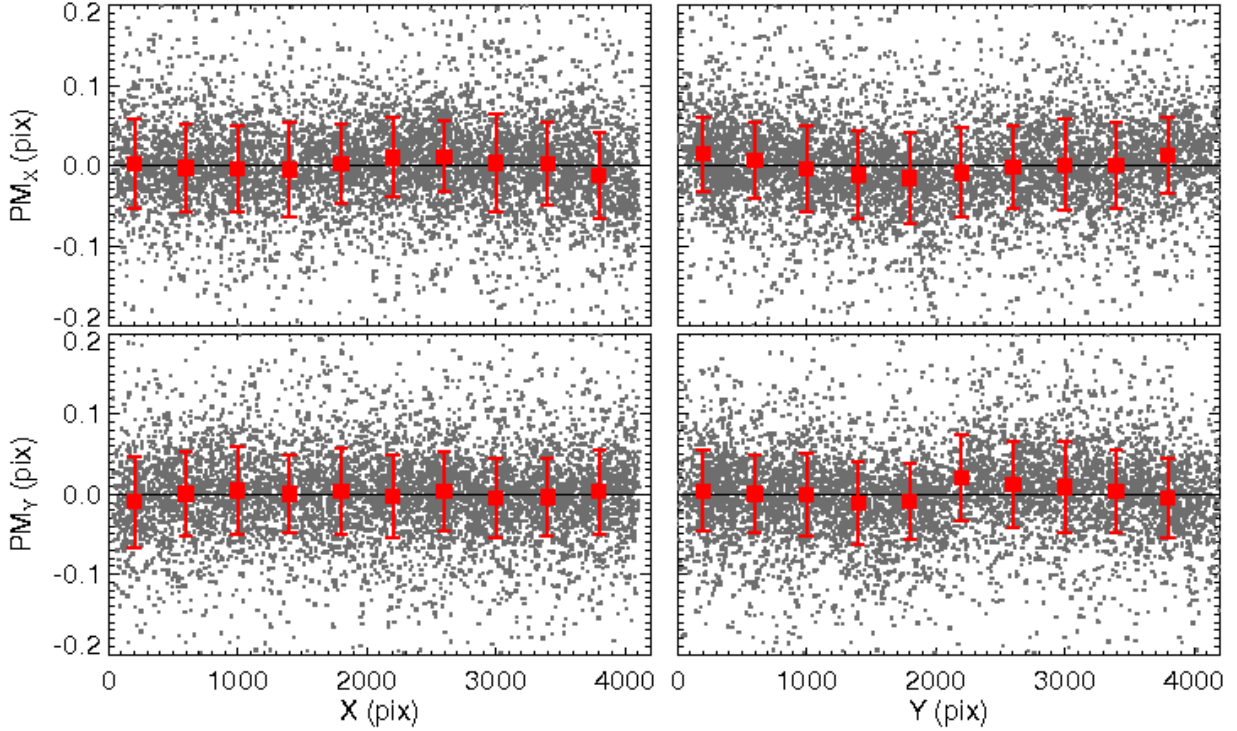}
\caption{Displacements of individual stars (dark gray dots) versus
         detector location between one of the second-epoch ACS/WFC
         images ({\tt jb4404vsq}) of the {\bf SPHEROID} field and the
         average of the first-epoch images, plotted separately for $X$
         and $Y$ positions. The units are in native ACS/WFC pixels,
         and $X$ and $Y$ positions are in the reference frame.  We
         also plot the average displacements of stars and the RMS of
         the distribution for each 400-pixel bin in red. The RMS is
         equivalent to the average uncertainty for an individual
         star. The 1-$\sigma$ error bars on the red data points equal
         RMS$/\sqrt{N}$, where $N$ is the number of stars in the bin,
         and these are smaller than the sizes of the points
         themselves. So while the displacements average to zero by
         construction, there are statistically significant low-level
         trends indicative of residual detector effects. We correct
         the measurements for these trends using the local corrections
         described in Section~\ref{sec:DerivingPMs}.
\label{fig:starpm1}}
\end{figure}

\begin{figure}
\epsscale{1.0}
\plotone{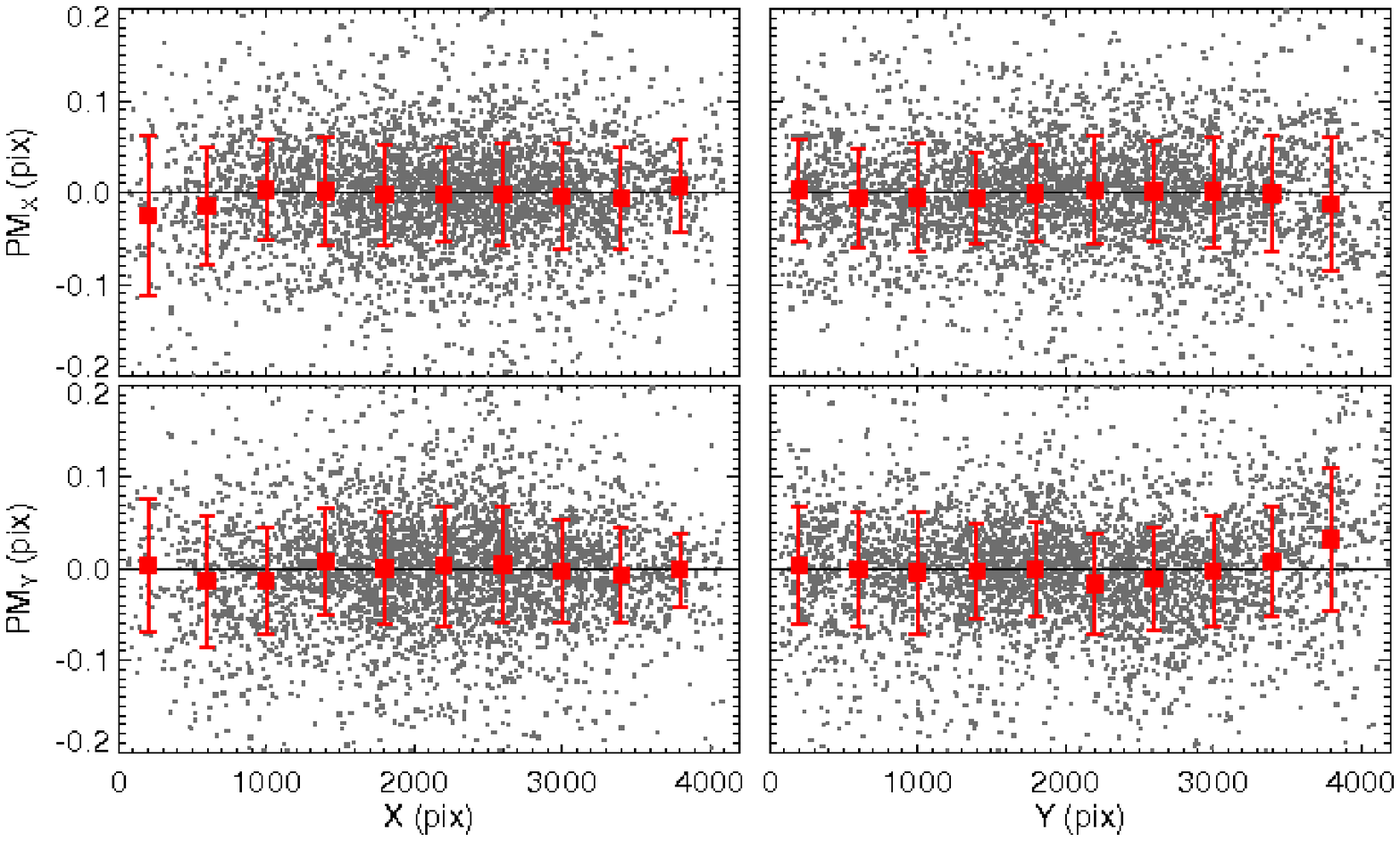}
\caption{As Figure~\ref{fig:starpm1}, but now for one of the 
         second-epoch WFC3/UVIS images ({\tt ib4401rsq}) of the 
         {\bf SPHEROID} field.
\label{fig:starpm2}}
\end{figure}

By construction, our method aligns the star fields between epochs. 
M31 stars will therefore have zero PM on average. 
Figures~\ref{fig:starpm1} and \ref{fig:starpm2} show the
residual motion of each star in $X$ and $Y$ ($PM_{X}$ and $PM_{Y}$) 
as a function of chip coordinates for one of the second-epoch ACS/WFC
images, and for one of the second-epoch WFC3/UVIS images. These plots
show that the ``average star'' (given our selection criteria and
brightness limits) is measured with a precision of $\sim 0.05$
pixel (but as shown in Figure~\ref{fig:plot_sigxy}, this is a strong 
function of brightness). When averaging over an ensemble of $N$ stars 
in $M$ exposures, the accuracy of the average PM is in principle 
smaller by a factor $\sqrt{NM}$. The stellar PMs show that there 
remain some small systematic trends with position on the detector, 
at levels $\lesssim 0.02$ pixel. These trends may be related to small 
changes in the distortion solution between the first and second 
epochs, or other low-level systematic effects that are not accounted 
for in our analysis. Such systematic trends might affect PM results. 
It is therefore important that we correct for them. We note that we 
found similar residual trends in all three target fields. This 
indicates that whatever is causing the low-level systematic trends 
is related to the detector characteristics rather than the internal 
kinematics of M31.

The background galaxies were not used to define any transformations
between the first-epoch and second-epoch data. Any motion between the
galaxies and stars should therefore show up as a displacement between
the first- and second-epoch galaxy positions in the reference
frame. We measure these displacements in a two step process to remove
any systematic PM trends related to the position on the detector, 
such as those visible in Figures~\ref{fig:starpm1} and 
\ref{fig:starpm2}.

First, we calculate the difference between the first- and 
second-epoch positions for each background galaxy in the reference 
frame. Then, for each galaxy, we compute the average displacement of 
stars in the vicinity of the galaxy. We subtract this displacement 
from the galaxy displacement. This ``local correction'' ensures that 
the displacement of each background galaxy is measured only with 
respect to the M31 stars that fall on the same part of the detector. 
This removes any remaining systematic PM residual associated with 
detector position. Each local correction is constructed using stars 
of similar brightness ($\pm 1$ mag in the inner 5$\times$5 pixels 
area) and within a 200 pixel region centered on the given background 
galaxy. The matching of the stars and background galaxies in 
brightness was motivated by the fact that some known detector effects 
(such as CTE), depend not only on detector position, but also on the 
source brightness.

\begin{figure}
\epsscale{1.0}
\plotone{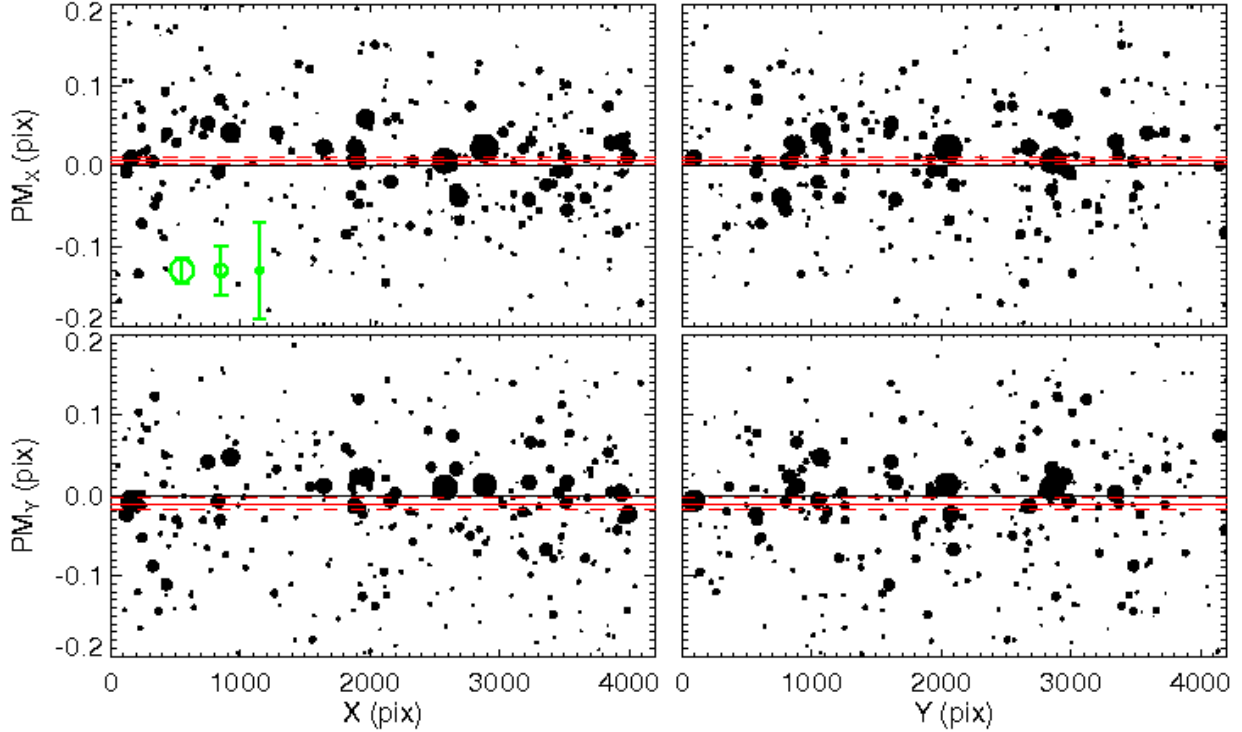}
\caption{Displacements of background galaxies versus detector 
         location between one of the second epoch ACS/WFC images 
         ({\tt jb4404vsq}) of the {\bf SPHEROID} field and the 
         average of the first-epoch images, plotted separately for 
         $X$ and $Y$ positions. The black points show the relative 
         displacements measured for different background galaxies. 
         The weighted average for all galaxies is shown as the red 
         line; dashed red lines indicate the 1-$\sigma$ error region 
         around the average. This region is smaller than the scatter 
         between the points by a factor of $\sim \sqrt{N}$, where $N$ 
         is the number of background galaxies. The radius of each 
         black point is proportional to $1/\Delta$, where $\Delta$ is 
         the PM measurement uncertainty for the particular background 
         galaxy. Hence, the area of each point is proportional to the 
         weight a point receives in the final weighted average. 
         Symbols in green in the top left panel illustrate how the 
         point size relates to the PM uncertainty $\Delta$. The units 
         are in native ACS/WFC pixels, and $X$ and $Y$ positions are 
         in the reference frame.
\label{fig:glxypm1}}
\end{figure}

\begin{figure}
\epsscale{1.0}
\plotone{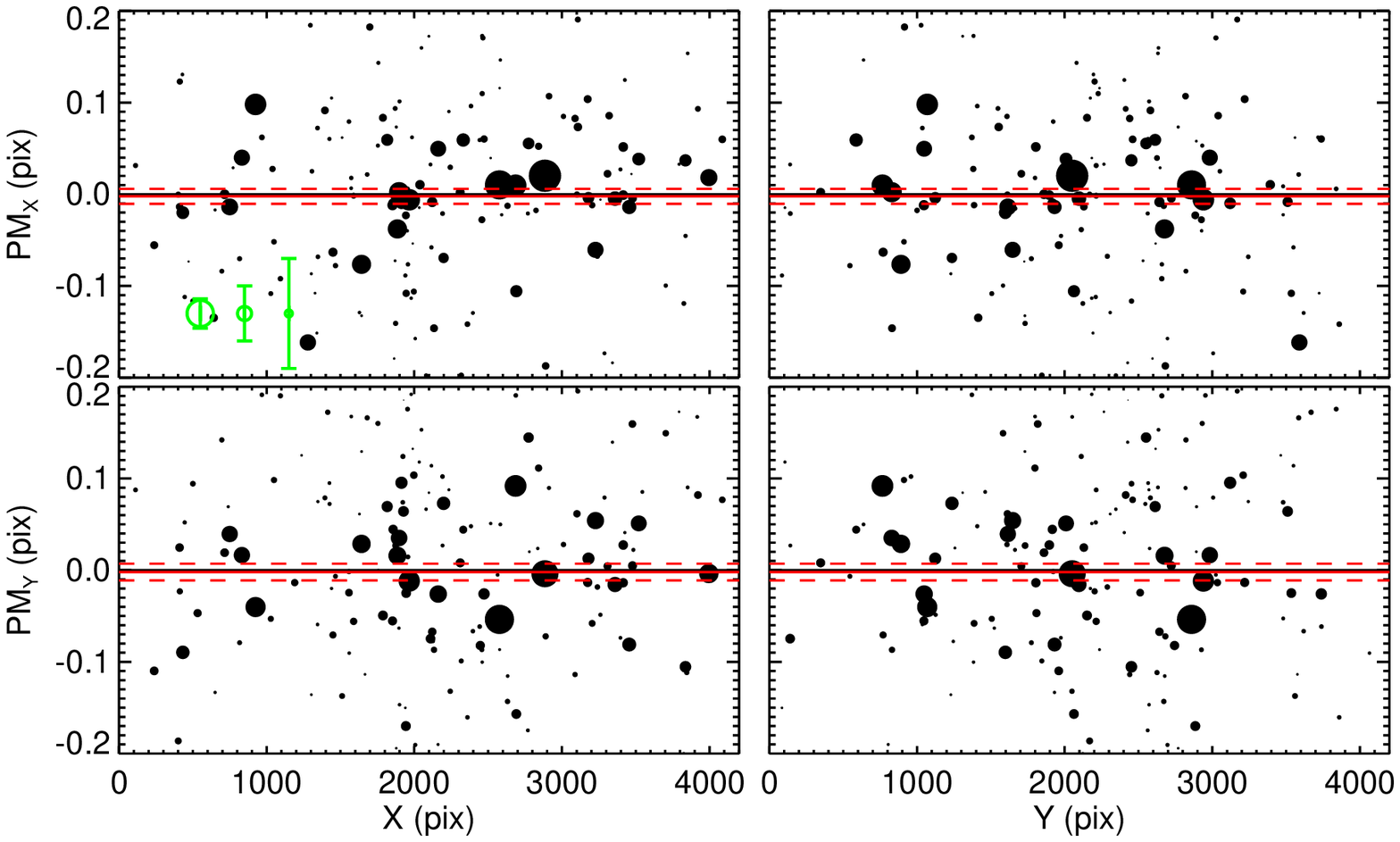}
\caption{As Figure~\ref{fig:glxypm1}, but now for one of the 
         second-epoch WFC3/UVIS images ({\tt ib4401rsq}) of the 
         {\bf SPHEROID} field.
\label{fig:glxypm2}}
\end{figure}

In the top panels of Figures~\ref{fig:glxypm1} and \ref{fig:glxypm2},
we show the $X$ and $Y$ proper motions ($PM_{X}$ and $PM_{Y}$) of 
each galaxy as a function of chip locations ($X$ and $Y$), for the 
same second-epoch exposures as in Figures~\ref{fig:starpm1} and
\ref{fig:starpm2}. There are many more stars than background galaxies 
in each frame, and the positions of the stars are generally more 
accurately determined than those of the galaxies (see 
Figure~\ref{fig:plot_sigxy}). The final PM error for each 
second-epoch exposure is therefore dominated entirely by the 
astrometric accuracy for the background galaxies.
  
\section{M31 Proper Motion}
\label{sec:M31PM}

\subsection{Final Results}
\label{sec:final}

To determine an average PM of the M31 stars in each field, we start
with the measured PM displacements of the background galaxies in the
reference frame (including local corrections as described above). For
each of the 18 different second-epoch images, we calculate the 
weighted average using the positional errors determined from the RMS 
among the first-epoch measurements. Outliers were rejected in this 
process with an iterative 3-$\sigma$ rejection scheme applied to the 
2-d PM distribution. The uncertainty of the average was computed 
using the bootstrap method \citep{efr93} with 10,000 bootstrap 
samples for each case. This provides a rigorous error estimate, which 
is ultimately based on the actual scatter between results inferred 
from different background galaxies. This naturally includes any 
random errors due to photon-counting statistics, as well as any 
potentially remaining systematic errors that lead to increased 
scatter (but any systematic errors that affect all sources equally 
would not be quantified by this).

\begin{figure}
\epsscale{1.0}
\includegraphics[scale=0.36]{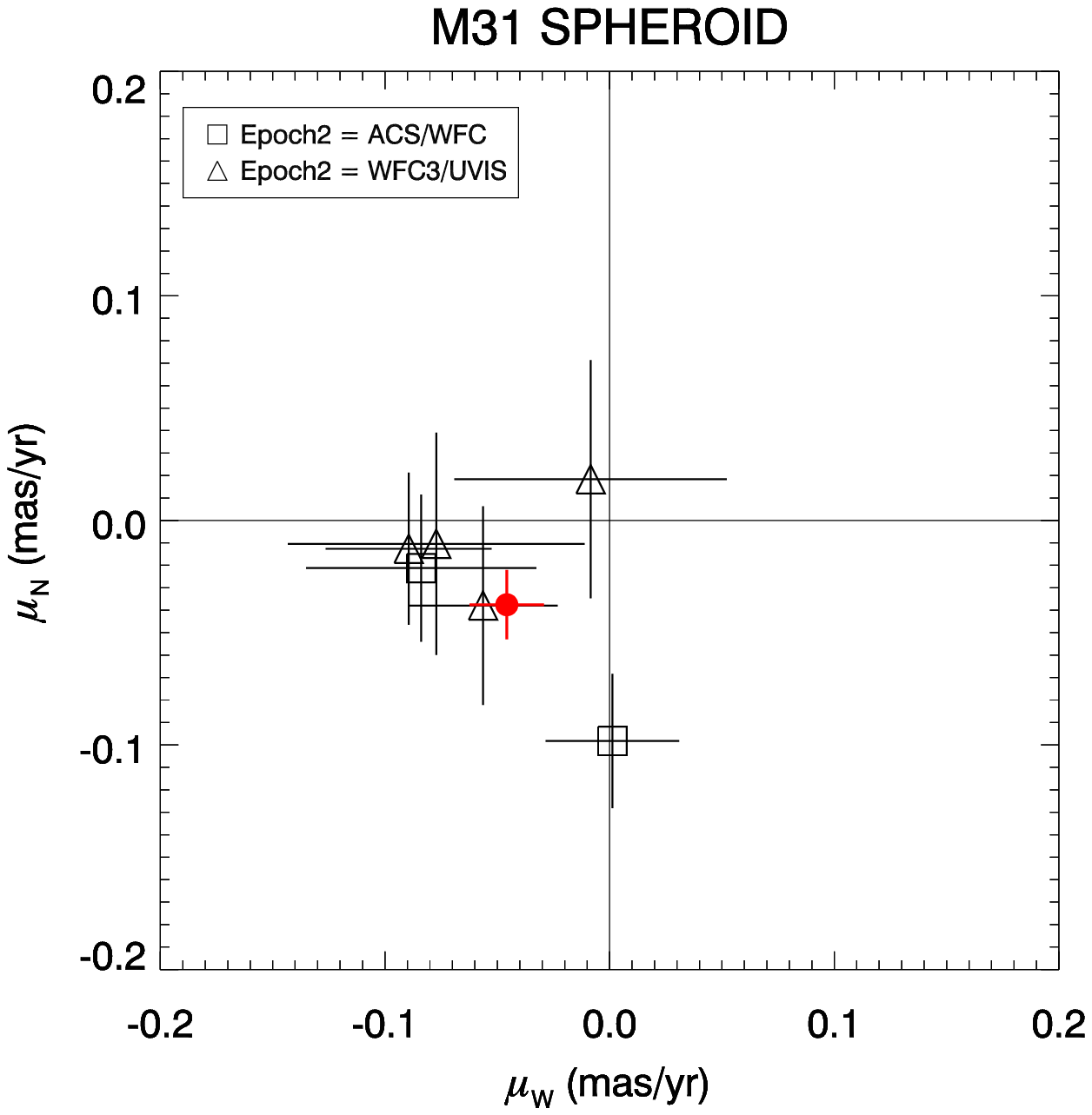}
\includegraphics[scale=0.36]{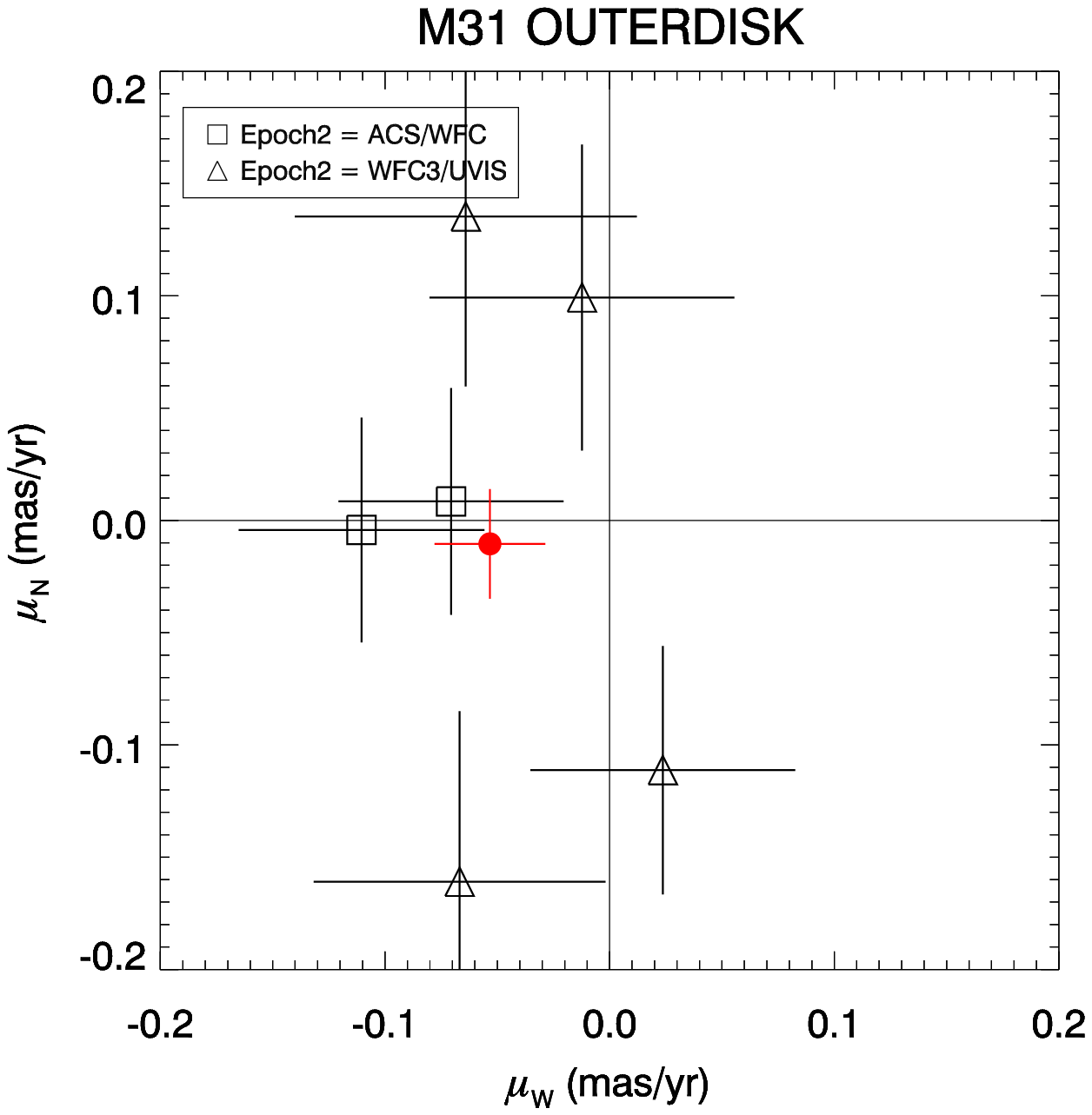}
\includegraphics[scale=0.36]{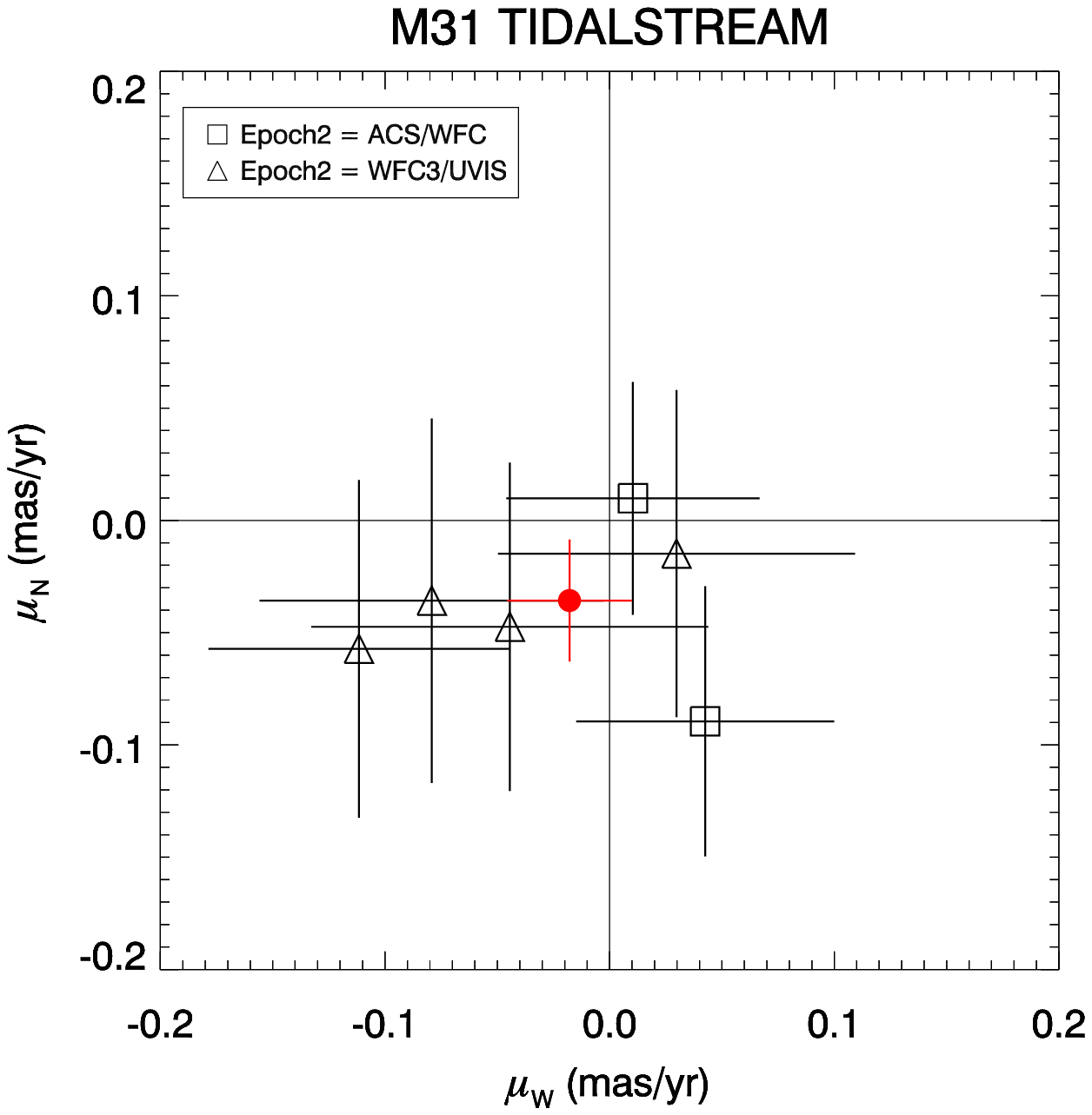}
\caption{Proper-motion results for the three target fields. Each 
         black symbol with an error bar indicates the weighted 
         average PM of M31 stars in the given field, inferred from a 
         single second-epoch exposure as in Figures~\ref{fig:glxypm1} 
         and \ref{fig:glxypm2}. Measurements using ACS/WFC (open 
         squares) and WFC3/UVIS (open triangles) are indicated with 
         different symbols. The solid red data point is the weighted 
         average of the six separate measurements for each field. 
\label{fig:PMDiagrams}}
\end{figure}

We then transformed the results to average PMs (and their associate
errors) of M31 stars in the directions north and west, in units of 
mas yr$^{-1}$. To do this, we used the orientation of the reference 
image (see Section~\ref{sec:FirstEpochACS}) with respect to the sky 
(using the FITS header keyword ORIENTAT), the time baseline, and the 
fact we are measuring the relative displacement of the background 
galaxies with respect to the M31 stars in our target fields. In 
Table~\ref{tab:Results1}, we tabulate the PM\footnote{$\mu_W$ is 
defined to be positive for an object moving towards the West, and 
therefore has the opposite sign of the PM in the RA direction.} 
$(\mu_W,\mu_N)$ inferred from each second-epoch image and the 
corresponding error, along with the number of background galaxies 
used for the PM derivation. Figure~\ref{fig:PMDiagrams} shows for 
each of the three target fields the PM estimates inferred from the 
six independent second-epoch exposures (2 ACS/WFC $+$ 4 WFC/UVIS).

The final estimate of the average PM of the M31 stars in each field 
is calculated by taking the error-weighted mean of the six 
independent measurements listed in Table~\ref{tab:Results1}. The 
results are shown in red in Figure~\ref{fig:PMDiagrams}, and they are 
tabulated in Table~\ref{tab:Results2}.

Figure~\ref{fig:PMAll} compares the final PM results for the three
different M31 fields. The weighted average of the results is shown in 
black. This weighted average is also listed in
Table~\ref{tab:Results2}. The figure shows results in physical units
of km s$^{-1}$, instead of mas yr$^{-1}$. To transform the units, we 
assumed a distance of 770 kpc \citep[see references in][]{vdm08}, 
so that $0.1$ mas yr$^{-1} = 365$ km s$^{-1}$. Distance errors were 
not propagated in this conversion. The final weighted average PM 
differs from zero at the $\sim 4.3$-$\sigma$ level. Therefore, we 
have actually measured a motion, and we have not merely put an upper 
limit on any motion.

\subsection{Consistency Checks}
\label{sec:checks}

To have faith in the results, it is important to assess the internal
consistency of the measurements. There are many checks available for
this, since we have performed measurements using different exposures, 
with different instruments, and for different fields.

For a given field $i$ ({\bf SPHEROID}, {\bf OUTERDISK}, or {\bf
TIDALSTREAM}), second-epoch instrument $j$ (ACS/WFC or WFC3/UVIS), 
and coordinate direction $k$ (West or North), we identify the set of 
$l$ PM measurements $\mu_{ijkl}$ (either four or two measurements,
depending on the instrument) with random errors $\Delta\mu_{ijkl}$. 
There are a total of 36 measurements (see Table~\ref{tab:Results1}). 
We combine the different measurements $l$ to obtain the 12 weighted 
averages ${\overline \mu}_{ijk}$ with random errors 
$\Delta {\overline \mu}_{ijk}$.

The quantity 
\begin{equation}
  \chi^2_{1} = \sum_{ijkl} 
     \left ( { { \mu_{ijkl} - {\overline \mu}_{ijk} } \over
               { \Delta \mu_{ijkl} } } \right )^2 
\end{equation}
provides a measure of the extent to which different measurements for
the same field and with the same instrument agree to within the 
random errors. We find that $\chi^2_{1} = 26.2$. In absence of 
systematic errors, one expects $\chi^2_{1}$ to follow a $\chi^2$ 
probability distribution with $N_{DF} = 36 - 12 = 24$ degrees of 
freedom. The expectation value for such a distribution is $N_{DF}$, 
and the dispersion is $\sim \sqrt{2 N_{DF}}$. Hence, the measurements 
for the same field with the same instrument are statistically 
consistent with each other. This is also visually evident from 
inspection of Figure~\ref{fig:PMDiagrams}, which shows furthermore 
that the agreement is least good for the {\bf OUTERDISK} field.

The quantity 
\begin{equation}
  \chi^2_{2} = \sum_{ik} 
             { { ({\overline \mu}_{i1k} - {\overline \mu}_{i2k})^2 } 
               \over
               { \Delta {\overline \mu}_{i1k}^2 + 
                 \Delta {\overline \mu}_{i2k}^2 } } 
\end{equation}
provides a measure of the extent to which measurements for the same
field with different instruments agree to within the random errors. 
We find that $\chi^2_{2} = 8.5$. In this case $N_{DF} = 6$, so the
measurements for the same field with the different instruments are 
also statistically consistent with each other.

Finally, the quantity 
\begin{equation}
  \chi^2_{3} = \sum_{ik} 
     \left ( { { {\overline \mu}_{ik} - {\overline \mu}_{k} } \over
               { {\Delta {\overline \mu_{ik}} } } } \right )^2 
\end{equation}
provides a measure of the extent to which measurements for different
fields agree to within the random errors. Here ${\overline \mu}_{ik}$
with random errors $\Delta {\overline \mu}_{ik}$ are the weighted
averages over all exposures for given field (top three lines of
Table~\ref{tab:Results2}). The ${\overline \mu}_{k}$ are the weighted
averages over all fields (bottom line of
Table~\ref{tab:Results2})). We find that $\chi^2_{3} = 1.9$. In this
case $N_{DF} = 6 - 2 = 4$, so the measurements for different fields
are also statistically consistent with each other. This is also
visually evident from inspection of Figure~\ref{fig:PMAll}.

The fact that all measurements are statistically consistent indicates
that there is no additional scatter in the data that is unaccounted
for by the random errors. This justifies the use of weighted averages
in combining the different measurements (which propagates only random
errors).

\subsection{Evaluating Systematic Errors}
\label{sec:syserr}

The final weighted average PM corresponds to a displacement of 0.0074
ACS/WFC pixel over the $\sim 7$ year time baseline for the {\bf
SPHEROID} field. The per-coordinate error in the displacement is only
0.0024 pixel. The displacement errors are even lower for the other 
fields, which have shorter time baselines.

These displacements are very small, but they can nonetheless be 
accurately measured. This can be understood based on first principles.
A single source can be centroided to an accuracy of order 
$\sigma/(S/N)$, where $\sigma$ is the Gaussian dispersion of the 
object size, and $S/N$ is the signal-to-noise ratio of the 
observation. For a compact source with $S/N \gtrsim 100$, this yields 
uncertainties of $0.01$--$0.02$ pixels, consistent with what we found 
in Figure~\ref{fig:plot_sigxy}. Therefore, even a single bright 
compact source in a single exposure can provide an accuracy 
comparable to the M31 PM displacement. Averaging over multiple 
sources in many exposures/fields reduces the random uncertainties to 
thousandths of a pixel and hence allows a solid measurement of the 
PM displacement.

The key to our robust measurement is, though, control of 
{\it systematic errors}, since those are not guaranteed to decrease 
by averaging multiple measurements. We have paid careful attention 
throughout our observation planning and subsequent analysis to 
identify possible sources of systematic errors and adjusted our 
methodology to minimize them as following:

\begin{itemize}

\item The observations were obtained with the F606W filter, 
which has the best PSF and distortion models for both ACS/WFC and
WFC3/UVIS (see Section~\ref{sec:Data2}).

\item The ACS/WFC images were obtained using the same telescope
orientation and pointing in both epochs (see Section~\ref{sec:Data2}). 
This allows any static astrometric residuals that depend on the 
position on the detector (e.g., static geometric distortion solution 
errors) to cancel out in the differential PM measurement.

\item The ACS/WFC data were explicitly corrected for the effects of
imperfect CTE, reducing any potential astrometric impact 
(see Section~\ref{sec:InitialProcessing}).

\item We used different geometric distortion solutions for the
different epochs of data, thus minimizing any impact due to potential 
time-variations in the geometric distortion solutions (see 
Section~\ref{sec:DistCor}).

\item We carefully selected our samples of M31 stars and background
galaxies to minimize any bias introduced by interlopers
(see Section~\ref{sec:StarGalaxyID}).

\item We built a unique template for every individual source that was
measured, thus avoiding any ad hoc assumptions about the properties 
of PSF or galaxy shapes (see Section~\ref{sec:TemplateFittingMethod}).

\item We fitted and accounted for PSF variations between different 
epochs, thereby minimizing any potential PM biases (see 
Section~\ref{sec:SecondEpochACS}).

\item We measured the relative PM of each background galaxy with 
respect to only its neighboring stars (the ``local correction''), 
thus minimizing the impact of spurious spatially varying PM residuals
(see Section~\ref{sec:DerivingPMs}).

\item The above local measurement was made with respect to only stars 
of similar brightness, thus minimizing the impact of 
magnitude-dependent PM residuals, including CTE effects (see 
Section~\ref{sec:DerivingPMs}).

\item We obtained the final PM for each field by averaging over
background galaxies at many different locations on the detector. 
This minimizes the potential impact of local peculiarities in 
detector properties or the astronomical scene (see
Section~\ref{sec:final}). Our approach is significantly advantageous  
over PM studies that use only a single background quasar in an image.

\item We determined the random PM error for each exposure from the
actual scatter between results from different background galaxies. 
The resulting errors include all sources of scatter, and not just 
those from photon-counting statistics (see Section~\ref{sec:final}).

\item We used robust statistical measures with outlier rejection
throughout our analysis, thus minimizing the potential influence of
individual erroneous measurements (e.g., Section~\ref{sec:final}).

\item We showed that PM measurements with different instruments, 
taken at different times and with different telescope orientations, 
as well as measurements of different fields, all yield statistically
consistent results (see Section~\ref{sec:checks}). This rules out a
large range of possible residual systematic errors, including most
possible astrometric residuals specific to a given instrument or
detector.

\end{itemize}

\noindent In summary, there is strong reason to believe that the PM
measurements and the quoted uncertainties are robust and free of
the potential systematics considered in this section. 

\subsection{M31 motions: Other Contributions and Constraints}
\label{sec:implications}

\begin{figure}
\epsscale{1.0}
\plotone{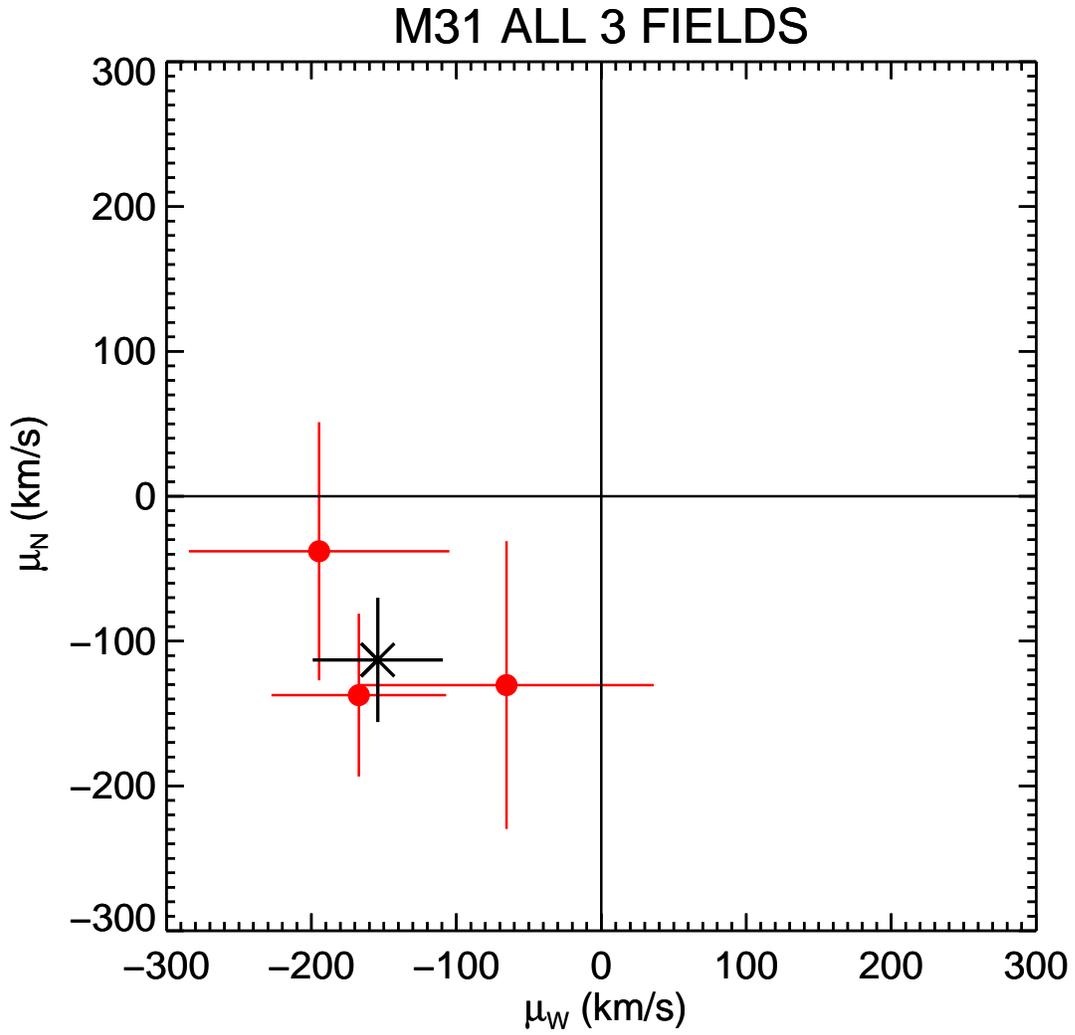}
\caption{Average proper motions, converted to km s$^{-1}$ using a
         distance to M31 of 770 kpc, for each target field (red 
         closed circles). The error-weighted mean of the 3 fields is 
         shown as the black {\tt X} mark. 
\label{fig:PMAll}}
\end{figure}

The results presented thus far do not directly measure the PM of the
M31 center of mass (COM). Instead, in every field we measure the sum
of the internal kinematics of the M31 stars and the COM motion. The
rotation curve of M31 has an amplitude of $\sim 250$ km s$^{-1}$  
\citep{cor10}. The contributions from internal kinematics are 
therefore not necessarily negligible compared to the uncertainties 
in our measurements (see Figure~\ref{fig:PMAll}).

In Paper~II we model the internal kinematics explicitly, and we 
derive an unbiased estimate for the COM PM. As it turns out, this 
estimate is not very different from the weighted average presented in
Table~\ref{tab:Results2}. We also show in Paper~II that this estimate
is statistically consistent with the estimate for the transverse
motion of M31 derived with the independent methods of \citet{vdm08},
based on the kinematics of M31 and Local Group {\it satellites}. This 
is a further indication that any remaining systematic uncertainties 
in our measurements are likely to be small.

The observed PMs presented here are heliocentric motions, i.e.,
{\it not} corrected for the reflex solar motion in the Milky Way. 
This known reflex motion falls in the same quadrant of PM space as 
our measurements \citep[see][]{vdm08}. The actual transverse motion 
of M31 with respect to the Milky Way is therefore closer to zero than 
what we show in Figure~\ref{fig:PMAll}. This provides another useful 
consistency check on our measurements, since there are theoretical 
reasons to suspect that the transverse motion of M31 with respect to 
the Milky Way should be small \citep[e.g.,][]{pee01}. In fact, we 
show in Paper~II that the transverse motion implied by our data is 
consistent with zero, meaning that M31 is likely moving on a nearly 
direct radial orbit towards the Milky Way.
 
\section{Conclusions}
\label{sec:conc}

We have presented the first direct absolute PM measurements of three 
fields in M31, our nearest giant companion galaxy in the Local Group. 
We used new second-epoch \hst\ data obtained with two different 
instruments, combined with very deep pre-existing first-epoch data, 
spanning a time baseline of 5--7 years. With state-of-the-art 
analysis methods, using background galaxies as stationary reference 
frame, we achieved a final PM accuracy of $\sim 12\ {\mu}$as 
yr$^{-1}$. This is comparable to the accuracies that have been 
achieved for other Local Group galaxies using VLBA observations of 
water masers \citep{bru05,bru07}. Water masers were recently 
discovered in M31 \citep{dar11}, but have yet to be used for a PM 
determination. We have paid careful attention to control of 
systematic errors throughout our analysis. A large range of 
consistency checks indicates that our PM measurements and the quoted 
uncertainties are robust and free of unknown systematics. The new PM 
results provide improved insights into the history, future, and mass 
of the Local Group. These topics are explored in detail in Papers~II 
and~III of this series.

The techniques presented here for measuring the PM of stars with
respect to background galaxies are not only applicable to M31, but
can be applied to a wide range of other problems. For example, our
group has ongoing \hst\ observing programs to use the same techniques
to measure the PM of Leo I (GO-12270, PI: S. T. Sohn), the PM of 
dwarf galaxies near the Local Group turn-around radius (GO-12273, 
PI: R. P. van der Marel), and the PM of stars in the Sagittarius 
Stream (GO-12564, PI: R. P. van der Marel). More generally, any deep 
wide-field space-based imager with a stable configuration, including 
future missions such as WFIRST and EUCLID, might be able to use the 
techniques presented here to measure the PMs of foreground objects 
from multiple epochs of data.

For M31 itself, it would be possible to improve the results presented 
here through additional \hst\ observations by, e.g., increasing the 
available number of fields, time baselines, or signal-to-noise ratio 
per epoch. Meanwhile, water masers hold the potential to soon provide 
PM measurements for individual sources in M31 \citep{dar11}. They 
will yield constraints on the transverse motion of the M31 
center-of-mass that are independent from those presented here and in 
Paper~II. Furthermore, water masers might allow measurement of 
additional effects, such as the M31 PM rotation, and the increase in 
M31's apparent size caused by its motion towards us.

\acknowledgments

Support for this work was provided by NASA through a grant for 
program GO-11684 from the Space Telescope Science Institute (STScI), 
which is operated by the Association of Universities for Research in 
Astronomy (AURA), Inc., under NASA contract NAS5-26555. The authors 
are grateful to Rachael Beaton, Gurtina Besla, Tom Brown, T. J. Cox, 
Mark Fardal, and Raja Guhathakurta for contributing to the other 
papers in this series, and for comments that helped improve the 
presentation of the present paper.

{\it Facilities:} \facility{HST (ACS/WFC; WFC3/UVIS)}.

%%%%%%%%%%%%%%%%%%%%%%%%%%%%%%%%%%%%%%%%%%%%%%%%%%%%%%%%%%%%%%%%%%%%%%
%% FIGURE 1
%%%%%%%%%%%%%%%%%%%%%%%%%%%%%%%%%%%%%%%%%%%%%%%%%%%%%%%%%%%%%%%%%%%%%%

\clearpage

%%%%%%%%%%%%%%%%%%%%%%%%%%%%%%%%%%%%%%%%%%%%%%%%%%%%%%%%%%%%%%%%%%%%%%
%% FIGURE 2
%%%%%%%%%%%%%%%%%%%%%%%%%%%%%%%%%%%%%%%%%%%%%%%%%%%%%%%%%%%%%%%%%%%%%%

\clearpage

%%%%%%%%%%%%%%%%%%%%%%%%%%%%%%%%%%%%%%%%%%%%%%%%%%%%%%%%%%%%%%%%%%%%%%
%% FIGURE 3
%%%%%%%%%%%%%%%%%%%%%%%%%%%%%%%%%%%%%%%%%%%%%%%%%%%%%%%%%%%%%%%%%%%%%%

\clearpage

%%%%%%%%%%%%%%%%%%%%%%%%%%%%%%%%%%%%%%%%%%%%%%%%%%%%%%%%%%%%%%%%%%%%%%
%% FIGURE 4
%%%%%%%%%%%%%%%%%%%%%%%%%%%%%%%%%%%%%%%%%%%%%%%%%%%%%%%%%%%%%%%%%%%%%%

\clearpage

%%%%%%%%%%%%%%%%%%%%%%%%%%%%%%%%%%%%%%%%%%%%%%%%%%%%%%%%%%%%%%%%%%%%%%
%% FIGURE 5
%%%%%%%%%%%%%%%%%%%%%%%%%%%%%%%%%%%%%%%%%%%%%%%%%%%%%%%%%%%%%%%%%%%%%%

\clearpage

%%%%%%%%%%%%%%%%%%%%%%%%%%%%%%%%%%%%%%%%%%%%%%%%%%%%%%%%%%%%%%%%%%%%%%
%% FIGURE 6
%%%%%%%%%%%%%%%%%%%%%%%%%%%%%%%%%%%%%%%%%%%%%%%%%%%%%%%%%%%%%%%%%%%%%%

\clearpage

%%%%%%%%%%%%%%%%%%%%%%%%%%%%%%%%%%%%%%%%%%%%%%%%%%%%%%%%%%%%%%%%%%%%%%
%% FIGURE 7
%%%%%%%%%%%%%%%%%%%%%%%%%%%%%%%%%%%%%%%%%%%%%%%%%%%%%%%%%%%%%%%%%%%%%%

\clearpage

%%%%%%%%%%%%%%%%%%%%%%%%%%%%%%%%%%%%%%%%%%%%%%%%%%%%%%%%%%%%%%%%%%%%%%
%% FIGURE 8
%%%%%%%%%%%%%%%%%%%%%%%%%%%%%%%%%%%%%%%%%%%%%%%%%%%%%%%%%%%%%%%%%%%%%%

\clearpage

%%%%%%%%%%%%%%%%%%%%%%%%%%%%%%%%%%%%%%%%%%%%%%%%%%%%%%%%%%%%%%%%%%%%%%
%% FIGURE 9
%%%%%%%%%%%%%%%%%%%%%%%%%%%%%%%%%%%%%%%%%%%%%%%%%%%%%%%%%%%%%%%%%%%%%%

\clearpage

%%%%%%%%%%%%%%%%%%%%%%%%%%%%%%%%%%%%%%%%%%%%%%%%%%%%%%%%%%%%%%%%%%%%%%
%% FIGURE 10
%%%%%%%%%%%%%%%%%%%%%%%%%%%%%%%%%%%%%%%%%%%%%%%%%%%%%%%%%%%%%%%%%%%%%%

\clearpage

%%%%%%%%%%%%%%%%%%%%%%%%%%%%%%%%%%%%%%%%%%%%%%%%%%%%%%%%%%%%%%%%%%%%%%
%% FIGURE 11
%%%%%%%%%%%%%%%%%%%%%%%%%%%%%%%%%%%%%%%%%%%%%%%%%%%%%%%%%%%%%%%%%%%%%%

\clearpage

%%%%%%%%%%%%%%%%%%%%%%%%%%%%%%%%%%%%%%%%%%%%%%%%%%%%%%%%%%%%%%%%%%%%%%
%% FIGURE 12
%%%%%%%%%%%%%%%%%%%%%%%%%%%%%%%%%%%%%%%%%%%%%%%%%%%%%%%%%%%%%%%%%%%%%%

\clearpage

%%%%%%%%%%%%%%%%%%%%%%%%%%%%%%%%%%%%%%%%%%%%%%%%%%%%%%%%%%%%%%%%%%%%%%
%% FIGURE 13
%%%%%%%%%%%%%%%%%%%%%%%%%%%%%%%%%%%%%%%%%%%%%%%%%%%%%%%%%%%%%%%%%%%%%%

\clearpage

%%%%%%%%%%%%%%%%%%%%%%%%%%%%%%%%%%%%%%%%%%%%%%%%%%%%%%%%%%%%%%%%%%%%%%
%% TABLE 1
%%%%%%%%%%%%%%%%%%%%%%%%%%%%%%%%%%%%%%%%%%%%%%%%%%%%%%%%%%%%%%%%%%%%%%

\clearpage
\begin{table}\scriptsize
\begin{center}
\caption{Description of Data for the Second Epoch Observations\label{tab:ObsLog}}
\vspace{3mm}
\begin{tabular}{l l c r@{.} l l r@{.} l}
\tableline\tableline
Target Field & Data set & Detector & \multicolumn{2}{c}{PA\_V3\tablenotemark{a}} & Exposure Time (s) & \multicolumn{2}{c}{$\Delta T$\tablenotemark{b}} \\
\tableline
\multirow{2}*{\bf SPHEROID}    & {\tt jb4404vsq, jb4404vuq}                       & ACS/WFC   &  75 & 07 & 1289, 1289             & 7 & 10     \\
                               & {\tt ib4401rsq, ib4401ruq, ib4401rxq, ib4401s1q} & WFC3/UVIS & 255 & 14 & 1379, 1379, 1450, 1450 & 7 & 57     \\
\tableline
\multirow{2}*{\bf OUTERDISK}   & {\tt jb4405jgq, jb4405jiq}                       & ACS/WFC   & 247 & 17 & 1289, 1289             & 5 & 06     \\
                               & {\tt ib4402urq, ib4402uwq, ib4402vqq, ib4402vuq} & WFC3/UVIS &  67 & 08 & 1420, 1420, 1491, 1491 & 5 & 55     \\
\tableline
\multirow{2}*{\bf TIDALSTREAM} & {\tt jb4406muq, jb4406mwq}                       & ACS/WFC   &  21 & 85 & 1299, 1299             & 5 & 99     \\
                               & {\tt ib4403enq, ib4403epq, ib4403esq, ib4403ewq} & WFC3/UVIS & 216 & 92 & 1379, 1379, 1450, 1450 & 5 & 47     \\
\tableline
\end{tabular}

\tablenotetext{a}{The position angle of the \hst\ V3 axis at the 
                  center of detector's field of view.}
\tablenotetext{b}{Baseline between the first and second-epoch data in 
                  years.}
\tablecomments{All data in the second epoch were obtained with the 
               F606W filter. Field coordinates and descriptions of 
               the first epoch-data are presented in \citet{bro06}.}
\end{center}
\end{table}

%%%%%%%%%%%%%%%%%%%%%%%%%%%%%%%%%%%%%%%%%%%%%%%%%%%%%%%%%%%%%%%%%%%%%%
%% TABLE 2
%%%%%%%%%%%%%%%%%%%%%%%%%%%%%%%%%%%%%%%%%%%%%%%%%%%%%%%%%%%%%%%%%%%%%%

\clearpage
\begin{table}\footnotesize
\begin{center}
\caption{Proper Motion Results for Individual Second Epoch Images\label{tab:Results1}}
\vspace{3mm}
\begin{tabular}{cl r@{.}l r@{.}l r@{.}l r@{.}l c}
\tableline\tableline
      &          & \multicolumn{2}{c}{$\mu_{W}$} & \multicolumn{2}{c}{$\mu_{N}$} & \multicolumn{2}{c}{$\sigma_{\mu_{W}}$} & \multicolumn{2}{c}{$\sigma_{\mu_{N}}$} &  \\
Field & Data Set & \multicolumn{2}{c}{(mas yr$^{-1}$)} & \multicolumn{2}{c}{(mas yr$^{-1}$)} & \multicolumn{2}{c}{(mas yr$^{-1}$)} & \multicolumn{2}{c}{(mas yr$^{-1}$)}    & $N_{\rm used}$\tablenotemark{a} \\
\tableline
                  & {\tt jb4404vsq} & $-0$ & 0839 & $-0$ & 0212 & 0 & 0512 & 0 & 0327 & 308 \\
                  & {\tt jb4404vuq} & $ 0$ & 0012 & $-0$ & 0982 & 0 & 0297 & 0 & 0299 & 306 \\
{\bf SPHEROID}    & {\tt ib4401rsq} & $-0$ & 0085 & $ 0$ & 0184 & 0 & 0606 & 0 & 0530 & 176 \\
                  & {\tt ib4401ruq} & $-0$ & 0564 & $-0$ & 0379 & 0 & 0332 & 0 & 0442 & 176 \\
                  & {\tt ib4401rxq} & $-0$ & 0772 & $-0$ & 0104 & 0 & 0660 & 0 & 0495 & 186 \\
                  & {\tt ib4401s1q} & $-0$ & 0895 & $-0$ & 0126 & 0 & 0369 & 0 & 0339 & 172 \\
\tableline
                  & {\tt jb4405jgq} & $-0$ & 0706 & $ 0$ & 0085 & 0 & 0501 & 0 & 0504 & 310 \\
                  & {\tt jb4405jiq} & $-0$ & 1105 & $-0$ & 0043 & 0 & 0547 & 0 & 0501 & 286 \\
{\bf OUTERDISK}   & {\tt ib4402urq} & $-0$ & 0123 & $ 0$ & 0993 & 0 & 0679 & 0 & 0681 & 156 \\
                  & {\tt ib4402uwq} & $-0$ & 0668 & $-0$ & 1609 & 0 & 0649 & 0 & 0759 & 152 \\
                  & {\tt ib4402vqq} & $-0$ & 0641 & $ 0$ & 1353 & 0 & 0761 & 0 & 0757 & 161 \\
                  & {\tt ib4402vuq} & $ 0$ & 0237 & $-0$ & 1112 & 0 & 0589 & 0 & 0553 & 152 \\
\tableline
                  & {\tt jb4406muq} & $ 0$ & 0425 & $-0$ & 0895 & 0 & 0574 & 0 & 0601 & 321 \\
                  & {\tt jb4406mwq} & $ 0$ & 0104 & $ 0$ & 0098 & 0 & 0564 & 0 & 0518 & 317 \\
{\bf TIDALSTREAM} & {\tt ib4403enq} & $-0$ & 1116 & $-0$ & 0572 & 0 & 0670 & 0 & 0752 & 185 \\
                  & {\tt ib4403epq} & $-0$ & 0792 & $-0$ & 0358 & 0 & 0766 & 0 & 0812 & 196 \\
                  & {\tt ib4403esq} & $-0$ & 0445 & $-0$ & 0474 & 0 & 0883 & 0 & 0731 & 175 \\
                  & {\tt ib4403ewq} & $ 0$ & 0298 & $-0$ & 0148 & 0 & 0794 & 0 & 0728 & 185 \\
\tableline
\end{tabular}
\tablenotetext{a}{Number of background galaxies used for deriving the 
                  average proper motion.}
\end{center}
\end{table}

%%%%%%%%%%%%%%%%%%%%%%%%%%%%%%%%%%%%%%%%%%%%%%%%%%%%%%%%%%%%%%%%%%%%%%
%% TABLE 3
%%%%%%%%%%%%%%%%%%%%%%%%%%%%%%%%%%%%%%%%%%%%%%%%%%%%%%%%%%%%%%%%%%%%%%

\clearpage
\begin{table}
\begin{center}
\caption{Final Proper Motion Results for the Three Target Fields\label{tab:Results2}}
\vspace{3mm}
\begin{tabular}{l r@{.}l r@{.}l r@{.}l r@{.}l}
\tableline\tableline
      & \multicolumn{2}{c}{$\mu_{W}$}       & \multicolumn{2}{c}{$\mu_{N}$}       & \multicolumn{2}{c}{$\sigma _{\mu_{W}}$} & \multicolumn{2}{c}{$\sigma_{\mu_{N}}$} \\
Field & \multicolumn{2}{c}{(mas yr$^{-1}$)} & \multicolumn{2}{c}{(mas yr$^{-1}$)} & \multicolumn{2}{c}{(mas yr$^{-1}$)}    & \multicolumn{2}{c}{(mas yr$^{-1}$)}    \\
\tableline
{\bf SPHEROID}    & $-0$ & 0458 & $-0$ & 0376 & 0 & 0165 & 0 & 0154 \\
{\bf OUTERDISK}   & $-0$ & 0533 & $-0$ & 0104 & 0 & 0246 & 0 & 0244 \\
{\bf TIDALSTREAM} & $-0$ & 0179 & $-0$ & 0357 & 0 & 0278 & 0 & 0272 \\
\tableline
weighted av.\tablenotemark{a} & $-0$ & 0422 & $-0$ & 0309 & 0 & 0123 & 0 & 0117 \\
\tableline
\end{tabular}
\tablenotetext{a}{Weighted average of the results for the 
                  three-target fields. This is {\it not} an unbiased 
                  estimate of the PM of the M31 center-of-mass. It 
                  contains contributions also from the internal 
                  motions of stars in M31, which are modeled and 
                  corrected in Paper~II.}
\end{center}
\end{table}


\clearpage
\begin{thebibliography}{M31PM}

\bibitem[Anderson \& King(2000)]{and00} Anderson, J., \& King, I. R.
    2000, \pasp, 112, 1360

\bibitem[Anderson(2005)]{and05} Anderson, J.  2005, in The 2005 HST 
    Calibration Workshop, ed. A. M. Koekemoer, P.Goudfrooij, 
    \& L. Dressel (Baltimore, MD: STScI)

\bibitem[Anderson \& King(2006)]{and06} Anderson J., \& King, I. R.
    2006, ACS/ISR 2006-01, PSFs, Photometry, and Astrometry for the 
    ACS/WFC (Baltimore: STScI) (AK06)

\bibitem[Anderson(2007)]{and07} Anderson, J.  2007, ACS/ISR 2007-08, 
    Variation of the Distortion Solution of the WFC (Baltimore: STScI)

\bibitem[Anderson \& van der Marel(2010)]{andvdm10} Anderson, J., \&
    van der Marel, R. P.  2010, \apj, 710, 1032

\bibitem[Anderson \& Bedin(2010)]{and10} Anderson, J. \& Bedin, L. R.
    2010, \pasp, 122, 1035

\bibitem[Baggett et al.(2011)]{bag11} Baggett, S., Bushouse, R., 
    Gilliland, R., Khozurina-Platais, V., Noeske, K., \& Petro, L. 
    2011, in WFC3 UVIS CTE Whitepaper,
    \url{http://www.stsci.edu/hst/wfc3/ins\_performance/CTE/cte.pdf}

\bibitem[Barnard(1917)]{bar17} Barnard, E. E.  1917, \aj, 30, 175

\bibitem[Bellini et al.(2011)]{bel11} Bellini, A., Anderson, J. \& 
    Bedin, L. R. 2011, \pasp, 123, 622

\bibitem[Bertin \& Arnouts(1996)]{ber96} Bertin, E., \& Arnouts, S.
    1996, \aaps, 117, 393

\bibitem[Brown et al.(2006)]{bro06} Brown, T. M., Smith, E., 
    Ferguson, H. C., Rich, R. M., Guhathakurta, P., Renzini, A., 
    Sweigart, A. V., \& Kimble, R.A. 2006, \apj, 652, 323

\bibitem[Brown et al.(2010)]{bro10} Brown, W. R., Anderson, J., 
    Gnedin, O. Y., Bond, H. E., Geller, M. J., Kenyon, S. J., \& 
    Livio, M. 2010, \apjl, 719, L23

\bibitem[Brunthaler et al.(2005)]{bru05} Brunthaler, A., Reid, M. J., 
    Falcke, H., Greenhill, L. J., \& Henkel, C.  
    2005, Science, 307, 1440

\bibitem[Brunthaler et al.(2007)]{bru07} Brunthaler, A., Reid, M. J., 
    Falcke, H., Henkel, C., \& Menten K. M.  2007, \aap, 462, 101

\bibitem[Corbelli et al.(2010)]{cor10} Corbelli, E., Lorenzoni, S., 
    Walterbos, R., Braun, R., \& Thilker, D. 2010, \aap, 511, A89

\bibitem[Cox \& Loeb(2008)]{cox08} Cox, T. J., \& Loeb, A.
    2008, \mnras, 386, 461

\bibitem[Darling(2011)]{dar11} Darling, J.  2011, \apjl, 732, L2

\bibitem[Efron \& Tibshirani(1993)]{efr93} Efron, B., \& 
    Tibshirani, R. 1993, An Introduction to the Bootstrap 
    (Chapman \& Hall/CRC)

\bibitem[Ferguson et al.(2002)]{fer02} Ferguson, A. M. N., 
    Irwin, M. J., Ibata, R. A., Lewis, G. F., \& Tanvir, N. R.
    2002, \aj, 124, 1452

\bibitem[Freedman \& Madore(1990)]{fre90} Freedman, W. L., \& 
    Madore, B. F. 1990, \apj, 365, 186

\bibitem[Fruchter \& Hook(2002)]{fru02} Fruchter, A. S., \& 
    Hook, R. N. 2002, \pasp, 114, 144

\bibitem[Fruchter(2011)]{fru11} Fruchter, A. S. 2011, 
    \pasp, 123, 497

\bibitem[Kalirai et al.(2007)]{kal07} Kalirai, J. S., Anderson, J.,
    Richer, H. B., King, I. R., Brewer, J. P., Carraro, G., 
    Davis, S. D., Fahlman, G. G., Hansen, B. M. S., Hurley, J. R., 
    L\'{e}pine, S., Reitzel, D. B., Rich, R. M., Shara, M. M., \& 
    Stetson, P. B. 2007, \apjl, 657, L93

\bibitem[Kallivayalil et al.(2006a)]{kal06a} Kallivayalil N., 
    van der Marel, R. P., Alcock, C., Axelrod, T., Cook, K. H., 
    Drake, A. J., \& Geha, M.  2006a, \apj, 638, 772

\bibitem[Kallivayalil et al.(2006b)]{kal06b} Kallivayalil, N., 
    van der Marel, R. P., \& Alcock, C.  2006b, \apj, 652, 1213

\bibitem[Lasker et al.(2008)]{las08} Lasker B. M. et al.  
    2008, \aj, 136, 735

\bibitem[Mahmud \& Anderson(2008)]{mah08} Mahmud N., \& Anderson, J.
    2008, \pasp, 120, 907

\bibitem[Milone et al.(2006)]{mil06} Milone, A. P., Villanova, S.,
    Bedin, L. R., Piotto, G., Carraro, G., Anderson, J., King, I. R., 
    \& Zaggia, S.  2006, \aap, 456, 517

\bibitem[Peebles et al.(2001)]{pee01} Peebles, P. J. E., 
    Phelps, S. D., Shaya, E. J., \& Tully, R. B.  2001, \apj, 554, 104

\bibitem[Piatek et al.(2002)]{pia02} Piatek, S., Pryor, C.,
    Olszewski, E. W., Harris, H. C., Mateo, M., Minniti, D.,
    Monet, D. G., Morrison, H., \& Tinney, C. G. 2002, \aj, 124, 3198

\bibitem[Piatek et al.(2003)]{pia03} Piatek, S., Pryor, C.,
    Olszewski, E. W., Harris, H. C., Mateo, M., Minniti, D., \& 
    Tinney, C. G.  2003, \aj, 126, 2346

\bibitem[Piatek et al.(2005)]{pia05} Piatek, S., Pryor, C.,
    Bristow, P., Olszewski, E. W., Harris, H. C., Mateo, M., 
    Minniti, D., \& Tinney, C. G.  2005, \aj, 130, 95

\bibitem[Piatek et al.(2006)]{pia06} Piatek, S., Pryor, C.,
    Bristow, P., Olszewski, E. W., Harris, H. C., Mateo, M., 
    Minniti, D., \& Tinney, C. G.  2006, \aj, 131, 1445

\bibitem[Piatek et al.(2007)]{pia07} Piatek, S., Pryor, C.,
    Bristow, P., Olszewski, E. W., Harris, H. C., Mateo, M., 
    Minniti, D., \& Tinney, C. G.  2007, \aj, 133, 818

\bibitem[Piatek et al.(2008)]{pia08} Piatek, S., Pryor, C., \& 
    Olszewski, E. W.  2008, \aj, 135, 1024

\bibitem[Press et al.(1992)]{pre92} Press, W. H., Teukolsky, S. A., 
    Vetterling, W. T., \& Flannery, B. P.  1992, Numerical Recipes in 
    FORTRAN (Cambridge: Cambridge Univ. Press)

\bibitem[Sohn, Anderson, \& van der Marel(2010)]{soh10} Sohn, S. T., 
    Anderson, J., \& van der Marel, R. P. 2010, in 2010 Space 
    Telescope Science Institute Calibration Workshop - Hubble after 
    SM4. Preparing JWST, ed. S. Deustua, \& C. Oliveira 
    (Baltimore, MD: STScI)

\bibitem[van der Marel et al.(2007)]{vdm07} van der Marel, R. P., 
    Anderson, J., Cox, C., Khozurina-Platais, V., Lallo, M., \& 
    Nelan, E. 2007, ACS/ISR 2007-07, Calibration of ACS/WFC Absolute 
    Scale and Rotation for Use in Creation of a JWST Astrometric 
    Reference Field (Baltimore: STScI)  

\bibitem[van der Marel \& Guhathakurta(2008)]{vdm08} 
    van der Marel, R. P., \& Guhathakurta, P.  2008, \apj, 678, 187

\bibitem[van der Marel et al.(2012)]{vdm12} van der Marel, R. P., 
    Fardal, M., Besla, G., Beaton, R. L., Sohn, S. T., Anderson, J., 
    Brown, T., \& Guhathakurta, P. 2012, \apj, submitted

\bibitem[Vieira et al.(2010)]{vie10} Vieira, K., Girard, T. M., 
    van Altena, W. F., Zacharias, N., Casetti-Dinescu, D. I., 
    Korchagin, V. I., Platais, I., Monet, D. G., L\'{o}pez, C. E., 
    Herrera, D., \& Castillo, D. J. 2010, \aj, 140, 1934

\end{thebibliography}
\end{document}